\documentclass[acmsmall,screen,fonts=false]{acmart}
\AtBeginDocument{%
  }

\setcopyright{acmlicensed}
\copyrightyear{2025}
\acmYear{2025}
\acmDOI{XXXXXXX.XXXXXXX}

\acmJournal{JACM}
\acmVolume{37}
\acmNumber{4}
\acmArticle{111}
\acmMonth{8}

\usepackage{wasysym}
\usepackage{color,soul}
\usepackage{xcolor}
\usepackage{listings}
\usepackage{subcaption}
\usepackage{multirow}
\usepackage{changepage}
\usepackage{threeparttable}
\usepackage{booktabs}
\usepackage{tcolorbox} 
\usepackage{natbib}
\usepackage{fontawesome5} 
\usepackage{colortbl}
\usepackage{verbatim}
\usepackage{graphicx} 
\newtcolorbox{boxK}{
    top=2pt,
    bottom=2pt,
    left=2pt,
    right=2pt,
    boxrule = 0pt,
    toprule = 0pt, 
}

\definecolor{codeblue}{RGB}{86, 168, 245}   
\definecolor{codebrown}{RGB}{170, 73, 38}   
\definecolor{codepurple}{RGB}{136,136,198} 
\definecolor{codegreen}{RGB}{106,171,115}    
\definecolor{codeorange}{RGB}{207, 142, 109}  
\definecolor{codeyellow}{RGB}{235, 210, 0}

\lstdefinestyle{numbers2} {
    deleteemph={[1]}, 
    deleteemph={[2]}, 
    deleteemph={[3]},
    deleteemph={[4]},
    deleteemph={[5]},
    resetmarginsandnumbers=true, 
    language=Python, 
    basicstyle=\small\ttfamily ,
    commentstyle=\color{gray},           
    stringstyle=\color{codegreen},         
    moredelim=[s][\color{black}]{field.type}{},
    emph={[1]transform\_sales\_records, generate\_sales\_report},
    emphstyle={[1]\color{codeblue}},          
    emph={[2]def,return,False,try,if,not,for,in,except,as, raise, hasattr},            
    emphstyle={[2]\color{codeorange}},     
    emph={[3]str, Exception, ValueError, type, int}, 
    emphstyle={[3]\color{codepurple}},  
    emph={[4]index,by,ascending},            
    emphstyle={[4]\color{codebrown}},
    emph={[5]@pytest\.mark\.xfail},            
    emphstyle={[5]\color{codeyellow}},
    breaklines=true,
    showstringspaces=false,
    frame=single,
    numbers=left,
}

\definecolor{instr}{RGB}{0, 80, 160}    
\definecolor{code}{RGB}{180, 0, 40}     
\definecolor{output}{RGB}{0, 120, 60}    
\definecolor{example}{RGB}{100, 60, 160} 

\newtcolorbox{promptbox}[1][]{colback=gray!5!white,
  colframe=black!80,
  boxrule=0.6pt,
  arc=2mm,
  left=2mm,
  right=2mm,
  top=1mm,
  bottom=1mm,
  fontupper=\ttfamily\footnotesize ,
  #1}
  
\soulregister{\cite}7 
\soulregister{\citep}7 
\soulregister{\citet}7 
\soulregister{\ref}7 
\soulregister{\pageref}7 

\begin{document}

\title{\textsc{Fixturize}: Bridging the Fixture Gap in Test Generation}


\author{Chengyi Wang}
\email{cy.w@mail.sdu.edu.cn}
\orcid{0009-0000-4153-6774}
\authornotemark[1]
\affiliation{%
   \institution{Shandong University}
   \city{Qingdao}
   \state{Shandong}
   \country{China}
}

\author{Pengyu Xue}
\email{xuepengyu@mail.sdu.edu.cn}
\orcid{0009-0007-3395-9575}
\authornote{These authors contributed equally to this work.}
\affiliation{%
   \institution{Shandong University}
   \city{Qingdao}
   \state{Shandong}
   \country{China}
}

\author{Zhen Yang}
\email{zhenyang@sdu.edu.cn}
\orcid{0000-0003-0670-4538}
\affiliation{%
   \institution{Shandong University}
   \city{Qingdao}
   \state{Shandong}
   \country{China}
}
\authornote{Corresponding author.}
\author{Xiapu Luo}
\email{csxluo@comp.polyu.edu.hk}
\orcid{0000-0002-9082-3208}
\affiliation{%
   \institution{The Hong Kong Polytechnic University}
   \city{Hong Kong}
   \country{China}
}
\author{Yuxuan Zhang}
\email{yx.zhang@mail.sdu.edu.cn}
\orcid{0009-0007-4791-3770}
\affiliation{%
   \institution{Shandong University}
   \city{Qingdao}
   \state{Shandong}
   \country{China}
}

\author{Xiran Lyu}
\email{xiranlyu@mail.sdu.edu.cn}
\orcid{0009-0002-8861-8907}
\affiliation{%
   \institution{Shandong University}
   \city{Qingdao}
   \state{Shandong}
   \country{China}
}

\author{Yifei Pei}
\email{peiyifei@mail.sdu.edu.cn}
\orcid{0009-0005-1351-0565}
\affiliation{%
   \institution{Shandong University}
   \city{Qingdao}
   \state{Shandong}
   \country{China}
}

\author{Zonghan Jia}
\email{zonghj@mail.sdu.edu.cn}
\orcid{0009-0007-9626-0569}
\affiliation{%
   \institution{Shandong University}
   \city{Qingdao}
   \state{Shandong}
   \country{China}
}

\author{Yichen Sun}
\email{yichensun050117@gmail.com}
\orcid{0009-0006-2605-4033}
\affiliation{%
   \institution{Henan University}
   \city{Zhengzhou}
   \state{Henan}
   \country{China}
}

\author{Linhao Wu}
\email{wulinhao@mail.sdu.edu.cn}
\orcid{0009-0001-7624-156X}
\affiliation{%
   \institution{Shandong University}
   \city{Qingdao}
   \state{Shandong}
   \country{China}
}

\author{Kunwu Zheng}
\email{Xiaozhengsdu2022@mail.sdu.edu.cn}
\orcid{0009-0000-7047-1981}
\affiliation{%
   \institution{Shandong University}
   \city{Qingdao}
   \state{Shandong}
   \country{China}
}

\renewcommand{\shortauthors}{Xue et al.}

\begin{abstract}

Current Large Language Models (LLMs) have advanced automated unit test generation but face a critical limitation: they often neglect to construct the necessary test fixtures, which are the environmental setups required for a test to run. To bridge this gap, this paper proposes \textsc{Fixturize}, a diagnostic framework that proactively identifies fixture-dependent functions and synthesizes test fixtures accordingly through an iterative, feedback-driven process, thereby improving the quality of auto-generated test suites of existing approaches. For rigorous evaluation, the authors introduce FixtureEval, a dedicated benchmark comprising 600 curated functions across two Programming Languages (PLs), i.e., Python and Java, with explicit fixture dependency labels, enabling both the corresponding classification and generation tasks.

Empirical results demonstrate that \textsc{Fixturize} is highly effective, achieving 88.38\%--97.00\% accuracy across benchmarks in identifying the dependence of test fixtures and significantly enhancing the Suite PaSs rate (SuitePS) by 18.03\%--42.86\% on average across both PLs with the auto-generated fixtures. Owing to the maintenance of test fixtures, \textsc{Fixturize} further improves line/branch coverage when integrated with existing testing tools of both LLM-based and Search-based by 16.85\%/24.08\% and 31.54\%/119.66\% on average, respectively. 
The findings establish fixture awareness as an essential, missing component in modern auto-testing pipelines.
\end{abstract}



\received{20 February 2007}
\received[revised]{12 March 2009}
\received[accepted]{5 June 2009}

\maketitle

\section{Introduction}

Automated Unit Test Generation is a long-standing challenge in software engineering \cite{10.1145/3092703.3092709}, aiming to generate effective tests that uncover potential faults and achieve high structural coverage \cite{8367053}. A typical unit test follows the \textit{Arrange–Act–Assert} pattern \cite{10859187,Meszaros2007}, where the \textit{Arrange} phase prepares the preconditions for execution, the \textit{Act} phase performs the operation under test, and the \textit{Assert} phase verifies the outcomes. With the rapid development of automated software testing, significant progress has been achieved in automating the \textit{Act} and \textit{Assert} phases \cite{chen2024chatunitest,10.1145/3691620.3695501,10.1145/3748505}. Existing approaches can now synthesize semantically meaningful inputs \cite{lukasczyk2022pynguin}, generate high-quality assertions \cite{watson2020learning, 10329992}, and even infer the behavioral intent of functions \cite{10989033,liu2025typeaware}.   However, the \textit{Arrange} phase, which involves constructing the test fixture, remains largely unsolved. 
Regardless of traditional or LLM-based approaches, they treat test fixtures as trivial or implicitly available, assuming that once the code and parameters are known, the test can be executed \cite{pan2024aster,ryan2024codeaware}. In reality, this assumption breaks down when the function relies on non-trivial environmental states, thus significantly constraining the practical applicability of State-Of-The-Art (SOTA) test generation \cite{zhang2025unit,10.1145/3691620.3696020}.

\begin{figure}
    \centering
    \includegraphics[width=1\linewidth]{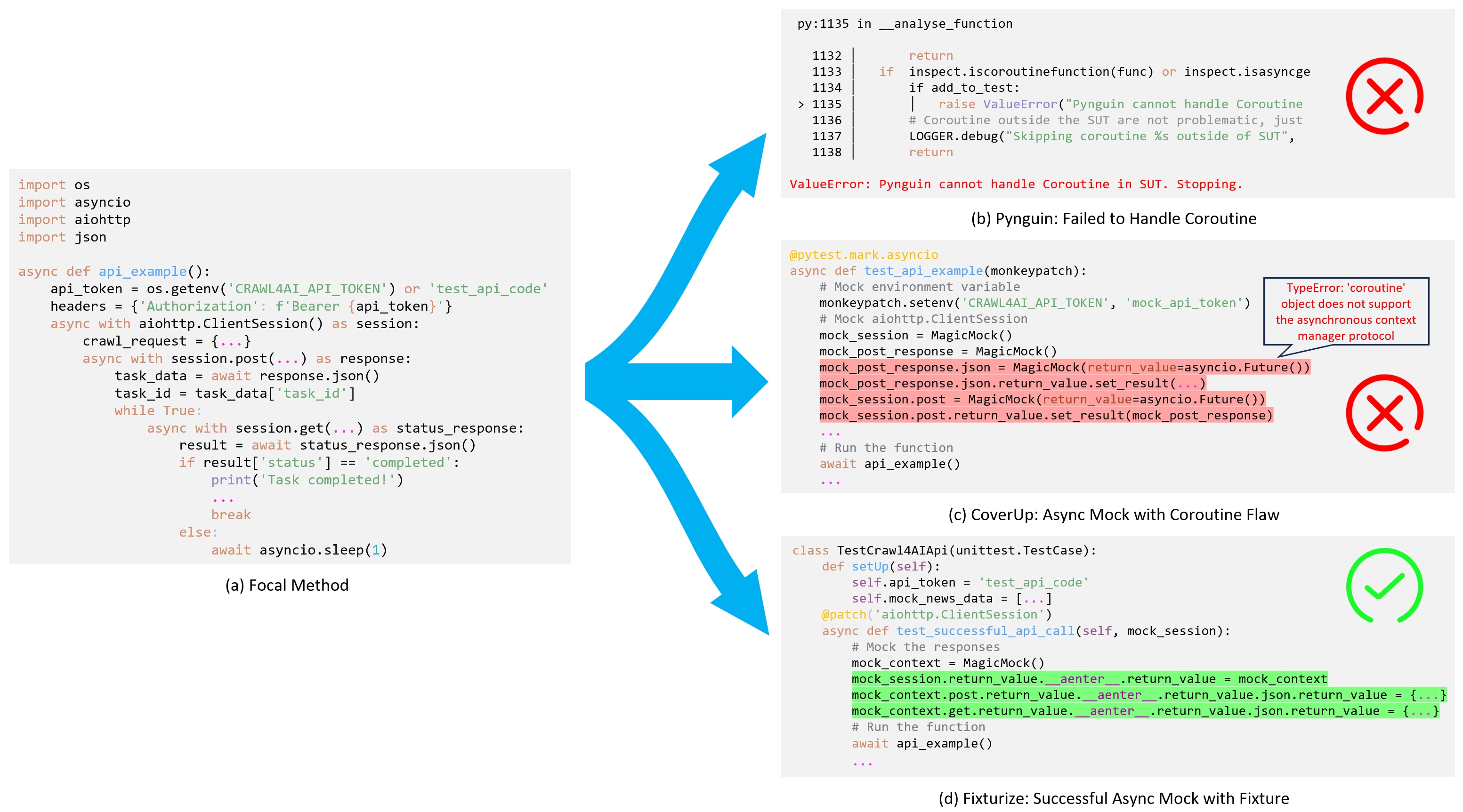}
    \caption{Illustration of the Fixture Gap and the Solution by \textsc{Fixturize}.}
    \label{fig:intro}
\end{figure}

To illustrate this limitation, consider the function \texttt{api\_example} which uses \texttt{aiohttp} to perform network requests. As shown in Figure \ref{fig:intro}, testing this function necessitates a specific execution context where the network session is properly mocked to simulate asynchronous behavior. We classify such functions as fixture-dependent. Traditional SBST tools such as \textsc{Pynguin} fail as they cannot handle Python coroutines and crash immediately upon encountering the keyword \texttt{async}. Besides, LLM-based tools such as C\textsc{over}U\textsc{p} attempt to generate a mock but struggle with the intricacies of the asynchronous context manager protocol. Specifically, C\textsc{over}U\textsc{p} incorrectly mocks \texttt{session.post} to return a \texttt{Future} object directly, failing to implement the required \texttt{\_\_aenter\_\_} and \texttt{\_\_aexit\_\_} methods. Consequently, when the code executes, it triggers a \texttt{TypeError} because the mock object does not support the asynchronous context manager protocol. 
An expected way in practice is: Firstly, identifying whether a test fixture is needed by \texttt{api\_example}. If needed, generate test fixtures to simulate the external dependency's behavior and state. Our proposed approach, Fixtureize, realizes this by correctly identifying the fixture dependency and synthesizing a \texttt{MagicMock} that explicitly implements the asynchronous protocol in the \texttt{setup} function as a test fixture, enabling successful execution. This stands in contrast to fixture-independent functions such as $\texttt{max(a, b)}$, which only require direct numerical parameters as inputs. Hence, a fundamental limitation persists in automated test generation: current SOTA algorithms, including both LLM-based and traditional techniques, cannot autonomously construct and manage these complex execution environments.

To address the above limitation, 
we propose \textsc{Fixturize}, an automated approach for fixture dependence identification and corresponding test generation.
It treats fixture construction as a proactive diagnostic process with a three-stage structured pipeline.
(1) The first stage identifies whether a focal method is fixture-dependent by evaluating whether it can be correctly invoked through a single-line function call without prior setup. If so, the focal method will be routed to direct test case generation. Otherwise, we deem it as fixture-dependent and ship it to the follow-up stages. A preliminary classification for fixture dependence is critical because generating unnecessary fixtures for fixture-independent test suites will probably induce extra unexpected mistakes without any benefits. (2) In the second stage,  \textsc{Fixturize} unlocks the restriction of single-line function calls and generates executable invocation examples for fixture-dependent functions through iterative refinement, guided by runtime feedback and adaptive hints.
(3) Finally, \textsc{Fixturize} harnesses the above invocation example to guide an LLM to synthesize test suites that incorporate complete fixtures, thereby ensuring the correctness of their test generation. 

For systematic evaluation of our framework, we construct FixtureEval, the first benchmark explicitly designed to measure fixture reasoning in test generation.  FixtureEval is composed of three subsets, namely $\text{FixtureEval}_{\text{G}}$, $\text{FixtureEval}_{\text{L}}$, and $\text{FixtureEval}_{\text{J}}$. $\text{FixtureEval}_{\text{G}}$ contains code samples written in Python drawn from popular GitHub repositories, while  $\text{FixtureEval}_{\text{L}}$ complements manually implemented leakage-free Python samples to prevent data contamination. Besides, code samples in $\text{FixtureEval}_{\text{J}}$ are also crawled from GitHub repositories, but are written in Java for cross-language examination. Each subset contains 200 code samples, where 100 samples are fixture-dependent, while the other 100 are fixture-independent, thereby enabling a dual evaluation on both classification and generation capability.
Experimental results demonstrate that \textsc{Fixturize} is highly effective in addressing the fixture gap. It achieves an 88.38\%--97.00\% accuracy in classifying fixture dependency, outperforming a direct LLM classification by 16.56\%--33.08\% on average across all subsets of FixtureEval. In terms of overall performance on FixtureEval, \textsc{Fixturize} significantly improves the Case PaSs rate (CasePS) by 10.71\%--15.98\%, Suite PaSs rate (SuitePS) by 18.03\%--42.86\% on average across both PLs against direct LLM prompting. More importantly, when integrated as a key diagnostic component with SOTA test generators, e.g., the LLM-based C\textsc{over}U\textsc{p} and the search-based \textsc{Pynguin}, \textsc{Fixturize} exhibits strong enabling capabilities. Across $\text{FixtureEval}_{\text{G}}$ and $\text{FixtureEval}_{\text{L}}$, it improves C\textsc{over}U\textsc{p}'s SuitePS on average by 8.51\% and further enhances its line and branch coverage by 16.85\% and 24.08\% owing to the maintenance test fixtures. As for the search-based \textsc{Pynguin}, \textsc{Fixturize} also on average boosts its line and branch coverage by 31.54\% and 119.66\%, respectively. 

In summary, this study makes the following contributions:
\begin{itemize}
    \item Introduction of \textsc{Fixturize}, an automated approach for fixture dependence identification and corresponding test generation, to bridge the fixture gap in existing test generation tools.
    \item Establishment of FixtureEval, the first dedicated benchmark for evaluating fixture dependence detection and fixture-dependent test generation across both statically- and dynamically-typed PLs, as well as a leakage-free version.
    \item Comprehensive evaluation across overall performance, ablation study, classification performance, integration with existing approaches, and cross-language generalization, empirically demonstrating the effectiveness of \textsc{Fixturize}.
\end{itemize}

\section{Related Work}
%
\subsection{Test Fixtures in Automated Testing}
Test fixtures define the controlled environment necessary for unit tests, including object initialization, dependency configuration, and the setup of external resources such as files, databases, or network services \cite{Meszaros2007,junitCookbook,pytestDoc}. Proper fixture design is critical for test reliability and maintainability, as complex or environment-dependent setups can lead to flaky or non-executable tests \cite{10.1145/3476105,TAHIR2023111837}. While modern testing frameworks provide structured setup and teardown mechanisms to standardize fixture handling \cite{nunitDoc,fraser2011evosuite}, automated test generation tools typically do not handle fixture construction themselves, implicitly assuming that the required fixtures already exist and lacking the capability to create or manage them \cite{dakhel2024effective,lemieux2023codamosa}. 
This limitation restricts the applicability of existing testing tools in practice and developers have to manually write fixture code before or after test generation, reducing the promised benefits of automation.

\subsection{Automated Test Generation}

Authoring high-quality tests is time-consuming. Daka et al. \cite{daka2014survey} revealed that developers, on average, spend 15\% of their time in test writing. Therefore, a wide range of automated testing tools, aiming to reduce the manual effort required for crafting test cases while uncovering potential software bugs \cite{KUMAR20168,10.5555/77517,Myers2012}, have been developed in recent years. 

\textbf{Traditional Approaches.}
Traditional automated approaches, such as Search-Based Software Testing (SBST) \cite{mcminn2004search, harman2015achievements}, symbolic execution \cite{king1976symbolic, cadar2008klee}, and fuzz testing \cite{zhu2022fuzzing,fioraldi2020afl++}, transform test generation into a testing input optimization problem towards improving testing coverage, mainly focusing on the \textit{Act} phase. For example, \textsc{EvoSuite} \cite{fraser2011evosuite} and \textsc{Pynguin} \cite{lukasczyk2022pynguin} are two typical SBST tools using evolutionary algorithms to automatically create test suites that maximize testing coverage. \textsc{KLEE} \cite{cadar2008klee} and \textsc{DART} \cite{godefroid2005dart} are automated test generation tools based on symbolic execution and constraint solving, but employ distinct strategies. The former reasons abstractly about all paths, while the latter adopts concrete runs with targeted symbolic analysis. In addition, \textsc{AFL++} \cite{fioraldi2020afl++}, as a community-powered fuzzing tool, operates on a genetic algorithm principle, where it automatically mutates input seeds and retains only those that trigger new code coverage in the focal methods, enabling it to efficiently explore deeper and more complex execution paths. However, their optimization algorithms are independent of the semantics of focal methods, making them incapable of handling complex environment setups (i.e., \textit{Arrange} phase) for running a test \cite{fraser2013evosuite,fraser2014large}. 



\textbf{Neural-based Approaches.}
Earlier works focus on the \textit{Assert} phase only, aiming to generate meaningful assert statements for test methods. For example, \textsc{ATLAS} \cite{watson2020learning} for the first time employs a Neural Machine Translation (NMT)-based approach for assertion generation. Follow-up studies, such as \cite{yu2022automated}, incorporate information retrieval into the model learning process for performance augmentation. With the development of deep learning, subsequent works gradually cover both the \textit{Act} and \textit{Assert} phases in the automated testing area, such as \textsc{AthenaTest} \cite{tufano2020unit}, \textsc{A3Test} \cite{alagarsamy2024a3test}, and \textsc{TeCo} \cite{nie2023learning}, started to utilize pre-training and supervised fine-tuning techniques to establish the test case generation capability of transformer models. Despite some success, training resources and model scales restrict their practical value compared with LLMs that carry powerful text-reasoning ability while bypassing fine-tuning. Consequently, numerous LLM-based test case generation approaches emerge \cite{lops2025system, nashid2023retrieval,dakhel2024effective,he2025hardtests}.  
\textsc{TestPolit} \cite{schafer2023adaptive} exploits the signature and implementation of focal methods, along with their usage examples extracted from documentation, to guide Codex to generate tests for npm packages. 
\textsc{ChatTester} \cite{Chattester}, \textsc{ChatUniTest} \cite{chen2024chatunitest}, and \textsc{TestForge} \cite{jain2025testforge} typically leverage iterative repair to augment the testing quality. \textsc{MuTAP} \cite{dakhel2024effective} exerts mutant testing to rectify unintended behaviors of test oracles in LLM-generated test cases. \textsc{TrickCatcher} \cite{liu2024llm} combined LLMs and differential testing to generate fault-revealing tests for plausible programs.
With the deepening of research, a series of works treat testing coverage as one of the most emphasized indicators and contribute substantial efforts.
\textsc{HITS} \cite{10.1145/3691620.3695501} employs program slicing to simplify focal methods, thereby increasing testing coverage. \textsc{IntUT} \cite{11029762} utilizes explicit branch-level test intentions to guide LLMs for coverage boosting. Besides, \textsc{SymPrompt} \cite{ryan2024code} and \textsc{CoverUp} \cite{altmayer2025coverup} both introduce static analysis to the LLM-powered test generation to improve testing coverage, but the latter's testing optimization is guided by coverage measurements, which is more direct and empirically effective.
Nonetheless, for setup failures that occur in the \textit{Arrange} phase, most approaches mentioned above treat them as generic errors, lacking proactive diagnostics and targeted fixing solutions. For example, one of the SOTA approaches, namely \textsc{CoverUp}, only fixes less than 40\% of fixture-related errors on average with its generic repair mechanism according to our experiments. Hence, generating high-quality tests for fixture-dependent functions remains a widespread blind spot for investigation till now.

\textbf{Hybrid Approaches.}
To fully combine the code generation ability of LLMs and testing concepts of traditional approaches, lots of hybrid approaches have emerged in recent years.
\textsc{CodaMosa} \cite{lemieux2023codamosa} improves SBST-based testing tools, e.g., \textsc{Pynguin}, by applying LLMs to generate tests for further mutating when the evolutionary algorithm reaches a fitness plateau. \textsc{TitanFuzz} \cite{deng2023large} makes the fuzzing protocol feasible in deep-learning framework testing via instructing LLMs to generate seed programs. In addition, they further proposed \textsc{Fuzz4All} to extend this idea to as many as nine software systems in six PLs for LLM-based fuzz testing \cite{xia2024fuzz4all}.
Although the aforementioned hybrid approaches are effective in their application areas, they follow the concepts of traditional testing tools, thus leaving the fixture gap still unresolved. 

\section{Approach}
In this section, we elaborate on the methodology and design motivations of \textsc{Fixturize}. 
\subsection{Overview}
\label{Overview}

\begin{figure}
    \centering
    \includegraphics[width=1\linewidth]{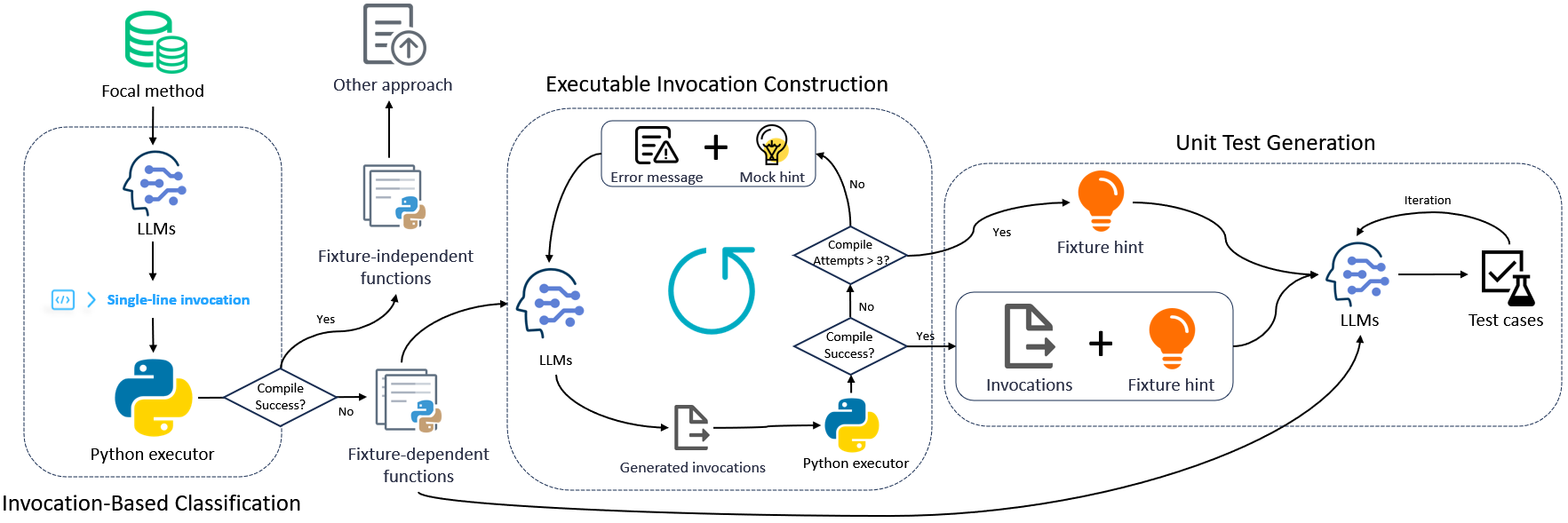}
    \caption{Workflow of \textsc{Fixturize}}
    \label{fig:workflow}
\end{figure}

Figure \ref{fig:workflow} presents an overview of \textsc{Fixturize}'s workflow, which accepts a focal method as input and outputs the corresponding test suite.
Overall, \textsc{Fixturize} contains three main components: 
(1) The first stage, namely Invocation-Based Classification (IBC), identifies whether a focal method is fixture-dependent by evaluating whether it can be correctly invoked through a single-line function call without prior setup. If so, the focal method is outside the scope of \textsc{Fixturize} and will be routed to other approaches for handling. Otherwise, \textsc{Fixturize} deems it as fixture-dependent and ships it to the follow-up stages. A preliminary classification for fixture dependence is critical because generating unnecessary fixtures for fixture-independent test suites will probably induce extra unexpected mistakes without any benefits. (2) In the second stage, i.e., Executable Invocation Construction (EIC), \textsc{Fixturize} unlocks the restriction of single-line calls and generates executable invocations for fixture-dependent functions through iterative refinement, guided by runtime feedback and adaptive hints. This is to provide usage exemplars for the subsequent test generation, such as necessary external dependencies and parameter configurations.
(3) Finally, in the Unit Test Generation (UTG) stage, \textsc{Fixturize} harnesses the above invocation example to guide an LLM to synthesize test suites that incorporate complete fixtures, thereby ensuring the correctness of their test generation.

\subsection{Step 1: Invocation-Based Classification (IBC)}
\label{Classifier}
The initial step of \textsc{Fixturize} involves categorizing functions as fixture-dependent or -independent, a crucial prerequisite for subsequent targeted test generation. However, accurately identifying fixture-dependent functions is inherently challenging. This difficulty stems from the fact that fixture dependency lacks deterministic surface patterns, which stands in stark contrast to more straightforward code properties, such as the use of third-party libraries. For these simpler properties, the classification task is relatively easy and can be reliably addressed using rule-based techniques such as regular expressions.
Figure \ref{fig:function-comparison} presents two examples to illustrate this challenge. As shown in panel (a), the function \verb|transform_sales_records| is fixture-independent, as it operates solely on its input parameters and requires no test setup beyond providing in-memory data. In contrast, panel (b) depicts the function \verb|generate_sales_report|, which is fixture-dependent because it must access a file specified by \verb|input_path|. This environmental constraint cannot be ascertained by static analysis alone. Notably, both functions import the same library and exhibit highly similar implementations, making rule-based discrimination between them infeasible. Therefore, distinguishing the dependence of test fixtures needs a semantic understanding of focal methods. 

\begin{figure}
    \centering
    \includegraphics[width=1\linewidth]{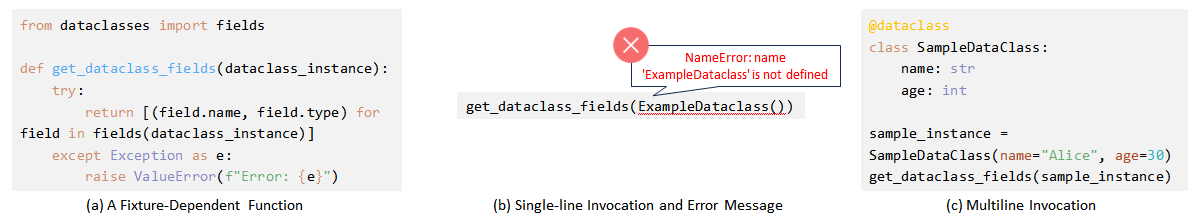}
    \caption{A Fixture-Dependent Function Cannot be Called Through Single-Line Invocations}
    \label{fig:single & multi}
\end{figure}

A straightforward workaround is feeding focal methods to LLMs for fixture dependence reasoning. However, as stated by Gendron et al. \cite{10.24963/ijcai.2024/693}, such abstract reasoning ability of current LLMs is flawed, making the direct output not convincing in theory. To improve the classification confidence, we reviewed a large number of code samples and found that the critical distinction between fixture-dependent and fixture-independent functions hinges on whether a function can be successfully invoked strictly through parameter passing within a single line, devoid of prior state setup. Hence, rather than dictating LLMs to directly respond a result, we instruct LLMs to generate a single-line invocation for the focal method and assume that if the generated single-line invocation is valid and runnable, the function tends to be fixture-independent; otherwise, it requires setup and is fixture-dependent. We use a simple example below to illustrate the so-called single-line invocation.
\begin{figure}[h]
    \centering
    \begin{subfigure}{0.95\textwidth}
    \centering
        \begin{lstlisting}[language=Python]
def transform_sales_records(sales_records: List[Dict[str, Any]]) -> List[Dict[str, Any]]:
    processed_df = pd.DataFrame(sales_records)
    #...
    result_df = processed_df.sort_values(by='total_value', ascending=False)
    return result_df.to_dict('records')
        \end{lstlisting}
        \caption{The Fixture-Independent Function}
    \end{subfigure}
    
    \medskip
    
    \begin{subfigure}{0.95\textwidth}
    \centering
        \begin{lstlisting}[language=Python]
def generate_sales_report(input_path: str, output_path: str) -> None:
    processed_df = pd.read_csv(input_path)
    #...
    result = processed_df.sort_values(by='total_value', ascending=False)
    result.to_csv(output_path, index=False)
        \end{lstlisting}
        \caption{The Fixture-Dependent Function}
    \end{subfigure}

    \caption{A Comparative Example of Fixture-Dependent and Fixture-Independent Functions}
    \label{fig:function-comparison}
\end{figure}
As depicted in Figure \ref{fig:single & multi}, a valid invocation of the function in panel (a) requires instantiating a \texttt{dataclass} object first and subsequently passing this instance as an argument. The code snippet in panel (b) is a standard single-line invocation. However, it fails as it cannot encapsulate the class definition and object instantiation together into the parameter passing. In contrast, the invocation in panel (c) is successful because the necessary preparation has been configured in advance, but violates the single-line invocation criterion. Thus, the focal method \texttt{get\_dataclass\_field} is classified as fixture-dependent in practice.

Building on this insight, we design an exclusive prompt for single-line invocation generation in Figure \ref{fig:classifier-prompts}, comprising four components:
(1) Task Instructions (Blue) that define the objective; (2) Code Context (Red) providing the focal method information; (3) Formatting Examples (Purple) to constrain output style; and (4) Output Constraints (Green) to ensure executability.
Crucially, the Output Constraints section defines a Python execution block where the generated single-line invocation is injected into a specific placeholder (i.e., ``\#the single-line function invocation example here''). The generated program is subsequently executed, allowing us to use the runtime return value as the definitive criterion for classification. As empirically validated in Section \ref{RQ3}, IBC provides a robust foundation for identifying fixture-dependent functions, ensuring the reliability of our subsequent generation pipeline.


\begin{figure}[htbp]
\centering

\begin{minipage}{\textwidth}
\begin{promptbox}[title=Prompt : Guiding the LLM to generate single-line invocations. ]
\textcolor{instr}{Generate a minimal one-line function invocation example based on the given Python code.}\\
\textcolor{code}{code: \{code\}}\\
\textcolor{example}{The output must follow the two examples:}\\
\textcolor{example}{1.Generate a minimal one-line function invocation example based on the given Python code.}\\
\textcolor{example}{ \ \ code:\{example code1\}}\\
\textcolor{example}{ \ \ output:\{single-line invocation1\}}\\
\textcolor{example}{2.Generate a minimal one-line function invocation example based on the given Python code.}\\
\textcolor{example}{ \ \ code:\{example code2\}}\\
\textcolor{example}{ \ \ output:\{single-line invocation2\}}\\
\textcolor{output}{Note that the output must be exactly one single-line function invocation, because it will be used in this framework:}\\
{\color{output}\verb|```python|}\\
 \textcolor{output}{if \_\_name\_\_ == "\_\_main\_\_":}\\
\textcolor{output}{ \ \ \ \ try:}\\
\textcolor{output}{ \ \ \ \ \ \ \ \ \#the single-line function invocation example here}\\
\textcolor{output}{ \ \ \ \ except Exception as e:}\\
\textcolor{output}{ \ \ \ \ \ \ \ \ print(e)}\\
\textcolor{output}{ \ \ \ \ \ \ \ \ exit(1)}\\
 {\color{output}\verb|```|}
\end{promptbox}
\end{minipage}

\caption{Prompt Used in the Classification Component}
\label{fig:classifier-prompts}
\end{figure}

\subsection{Step 2: Executable Invocation Construction (EIC)}
\label{Executable Invocation Construction}
Upon identifying fixture-dependent functions, the subsequent challenge lies in constructing the requisite execution environments. While directly generating test fixtures is complex, we observe that an executable function invocation already provides illustrations for a well-configured environment. For example, panel (c) in Figure \ref{fig:single & multi} presents an executable invocation for the fixture-dependent function \texttt{get\_dataclass\_fields}, and the required class, i.e., \texttt{SampleDataClass}, has been defined and instantiated for parameter passing. 
Driven by this insight, this step focuses on generating executable invocations, which are self-contained code snippets that encapsulate both the necessary setup logic (e.g., initializing objects, mocking services) and the function call itself, thereby acting as an exemplar for the follow-up test generation. 

\begin{figure}
    \centering
    \includegraphics[width=1\linewidth]{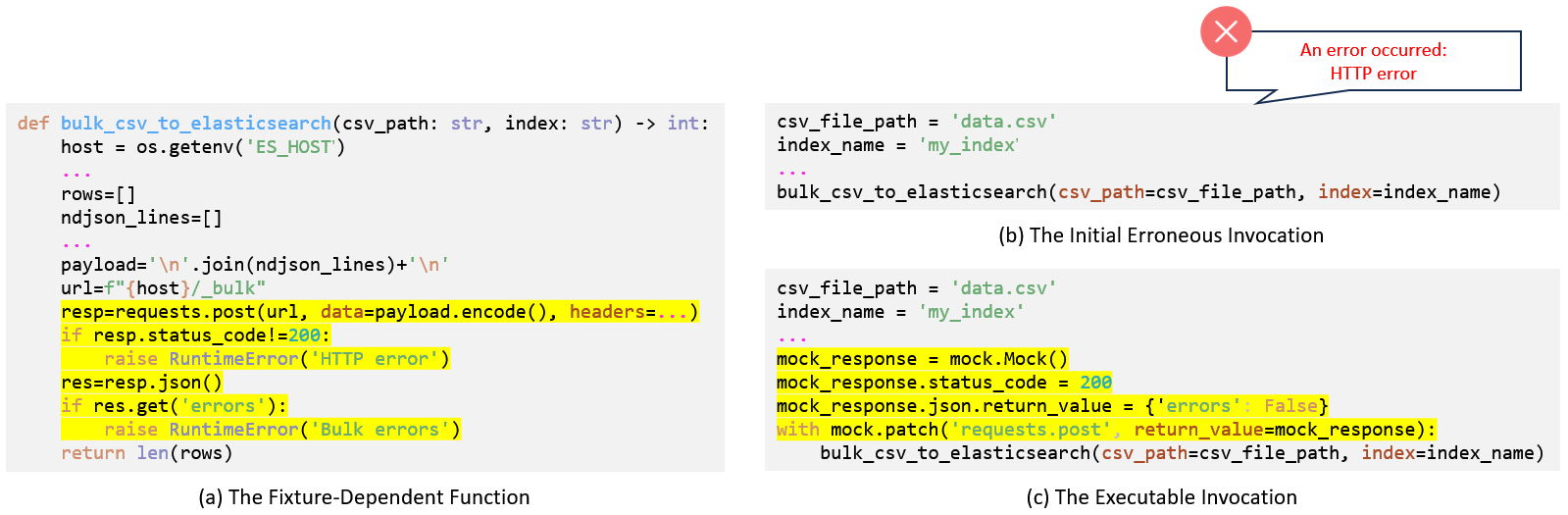}
    \caption{An Iterative Remediation Example of EIC }
    \label{fig:invocation}
\end{figure}

However, constructing such invocations presents a significant challenge when functions rely on external ecosystems. In practice, fixture-dependent functions often interact with external services that are absent in the isolated test environment, rendering direct execution unfeasible. Figure \ref{fig:invocation} illustrates this issue: the function's implementation is contingent upon an active Elasticsearch service \cite{Elasticsearch}. Without this external dependency, a standard invocation attempt inevitably leads to execution failure. To address this, we employ mocking to simulate these external interactions. Nevertheless, we apply this technique judiciously, as prior studies warn that indiscriminate mocking can lead to overfitting and test brittleness, thereby diminishing the suite's resilience to subsequent focal methods modifications \cite{9286134,7372009}. Balancing these trade-offs, we adopt the criteria established by Spadini et al. \cite{to_mock_or_not} to govern the injection of mocks. 
Specifically, \textsc{Fixturize} strategically restricts mocking to three categories: web links or services, databases, and external dependencies. 


\begin{figure}[htbp]
\centering
\begin{minipage}{\textwidth}
\begin{promptbox}[title=Prompt : Guiding the LLM to generate executable invocations. ]
\textit{\textbf{[Inital prompt:]}}\\
\textcolor{instr}{Generate an executable function invocation example for the following Python function. Given that the function is already provided before the invocation, do not show or import the function code.}

\textcolor{code}{Function code:}\textcolor{code}{\{original\_code\}}

\textit{\textbf{[If a previous attempt failed, append the following block:]}}\\
\textcolor{code}{Previous attempt (generated by you):}\\
\textcolor{code}{\{previous\_attempt\}}

\textcolor{code}{Execution result:}\\
\textcolor{code}{\{error\_message\}}

\textcolor{output}{If this error is caused by web link or services, database, or external dependencies, then introduce the mock method in the next instance generation. If not, there's no need for a mock. If necessary, use 'from unittest import mock' uniformly.}

\textcolor{instr}{Based on that, please generate a corrected invocation, and ensure all required imports and context are included.}
\end{promptbox}
\end{minipage}
\caption{Prompt Template for Execution Feedback}
\label{fig:prompt:invocation}
\end{figure}

To realize the above methodology, we design an iterative refinement paradigm driven by execution feedback, as operationalized by the prompt structure in Figure \ref{fig:prompt:invocation}:
(1) Initial Prompt: The prompt begins by providing the source function code and explicitly instructing the LLM to generate an executable function invocation example.
(2) Feedback Injection: If a previous attempt fails, the prompt is augmented with the previously generated code and its corresponding error message.
(3) Mock Hint: Crucially, to handle environmental constraints, we inject a conditional mock hint. This instructs the LLM to introduce \texttt{unittest.mock} if the error indicates an environmental dependency failure. This validate-and-fix loop repeats for a maximum of three iterations. If the LLM fails to converge on an executable invocation within this limit, the system terminates the process for the current function. 
Figure \ref{fig:invocation} presents examples of LLM-generated invocations for a fixture-dependent function \texttt{bulk\_csv\_to\_elasticsearch}. Panel (a) illustrates the fixture-dependent function. Panel (b) displays the initial erroneous invocation, whereas panel (c) shows the final corrected invocation. The code highlighted in yellow in panel (a) indicates the segment responsible for the failure of the initial invocation in panel (b), with the corresponding corrections presented in panels (c). Initially, the LLM attempts a direct invocation. Although it correctly handles file system prerequisites, it fails to account for the active Elasticsearch service, resulting in an HTTP connection error. Because this network-related failure meets our mocking criteria, 
the LLM successfully generates the corrected invocation using \verb|unittest.mock| to simulate the external service, as shown in panel (c).

\subsection{Step 3: Unit Test Generation (UTG)}

\label{Unit Test Generation}
Building on the outputs of EIC, this final step generates the complete unit test suite. To maximize generation quality, \textsc{Fixturize} employs an adaptive prompting strategy that dynamically configures the LLM's instructions based on whether an executable invocation was successfully generated in the previous step. Figure \ref{fig:prompt-comparison} illustrates this division. The blue segments denote shared instructions that ground the generation in the provided source code, the red segment injects the successfully generated invocation as a reference, and the green segments explicitly mandate fixture construction.
\begin{figure}[htbp]
\centering
\begin{minipage}{\textwidth}
\begin{promptbox}[title=Prompt 1: Generating test cases with an invocation exemplar. ]
\textcolor{instr}{Based on the following code, please use 'unittest' to generate a Python test suite that includes 5 test cases. Import each focal method from the specified file using the syntax: `from \{base\_name\} import <func1>, <func2>···'.}\\
\textcolor{instr}{code: \{code\}}\\
\textcolor{code}{Here is its invocation example: }\\
\textcolor{code}{\{invocation\_example\}}\\
\textcolor{output}{This can help you generate the test fixture section. }
\end{promptbox}
\end{minipage}

\vspace{0.5cm}

\begin{minipage}{\textwidth}
\begin{promptbox}[title=Prompt 2: Generating test cases without invocation exemplars. ]
\textcolor{instr}{Based on the following code, please use 'unittest' to generate a Python test suite that includes 5 test cases. Import each focal method from the specified file using the syntax: `from \{base\_name\} import <func1>, <func2>···'.}\\
\textcolor{instr}{code: \{code\}}\\
\textcolor{output}{Ensure the presence of the test fixture in the test suite.}
\end{promptbox}
\end{minipage}

\caption{Adaptive Prompting Strategy of \textsc{Fixturize}}
\label{fig:prompt-comparison}
\end{figure}

To ensure the generated tests are structural and readable, we adopt the \texttt{unittest} framework from the Python standard library \cite{Ammann_Offutt_2016}. Furthermore, we direct the LLM to generate five distinct test cases per function. This design aligns with the cov@5 configuration introduced in \textsc{TestEval} \cite{wang-etal-2025-testeval}, where empirical evidence indicates that five test cases achieve substantial improvements in both coverage and diversity, with diminishing returns observed beyond this threshold. Furthermore, to mitigate import hallucinations caused by incomplete context, as noted by Yang et al. \cite{yang2025}, we explicitly prescribe the use of a structured import format: \verb|from {base_name} import <func1>, <func2>···|. This constraint serves to reduce syntactic and path-related errors in the generated code. In addition to the base prompt for unit test generation, we provide differentiated instructions depending on whether the invocation of the function is successfully generated in EIC. Crucially, for functions with pre-verified invocations, Prompt 1 leverages them as concrete examples to guide fixture setup; for others, Prompt 2 explicitly directs the LLM to generate appropriate test fixtures from scratch. 

Beyond initial generation, we incorporate execution feedback (e.g., error messages) as an additional mechanism to iteratively refine the generated test cases. Such feedback-driven approaches have demonstrated strong effectiveness in recent approaches, including C\textsc{over}U\textsc{p} \cite{altmayer2025coverup}, TestART \cite{testart2024}, GenX \cite{genx2024}, Mokav \cite{mokav2025} and so on. As illustrated in Figure \ref{fig:execution feedback}, when a generated test fails, the system constructs a remedial prompt containing the original function code, the failing test, and the specific error message. The blue section clarifies the repair objective, the red section provides the error context, and the green section constrains the output format to facilitate automated extraction \cite{altmayer2025coverup}. To balance computational cost with generation quality, we limit this process to a single-iteration refinement. If the revised test still fails after this attempt, the system discards it and proceeds to the next candidate, effectively filtering out persistent failures while correcting transient errors. Through this carefully designed prompt structure, augmented with execution feedback, we effectively guide the LLM to generate high-quality and contextually appropriate test cases. 
\begin{figure}[htbp]
\centering
\begin{minipage}{\textwidth}
\begin{promptbox}[title=Prompt : Test case iterative refinement. ]
\textcolor{instr}{The following Python test code failed to run. Please analyze the error and regenerate the correct test code:}\\
\textcolor{code}{Original function code:}\\
\textcolor{code}{\{function\_code\}}\\
\textcolor{code}{Original test code:}\\
\textcolor{code}{\{test\_code\}}\\
\textcolor{code}{error message:}\\
\textcolor{code}{\{error\_message\}}\\
\textcolor{output}{Please return the revised complete Python test code directly, only the code itself, without any explanation or comments. Ensure that the code format is correct and can be run directly.}
\end{promptbox}
\end{minipage}
\caption{Prompt Template for Iterative Refinement}
\label{fig:execution feedback}
\end{figure}

\section{Benchmark: FixtureEval }
\label{FixtureEval}
Existing benchmarks for test case generation \cite{wang2025projecttest,lukasczyk2022pynguin} do not annotate or distinguish between fixture-dependent and -independent functions. However, this distinction is essential for a comprehensive evaluation of \textsc{Fixturize}. On one hand, \textsc{Fixturize} must correctly identify which functions rely on fixtures; on the other hand, only after isolating these fixture-dependent functions can we meaningfully assess its ability to generate the corresponding tests with fixtures. Since no existing benchmark can adequately evaluate \textsc{Fixturize}, we introduce \textbf{FixtureEval}, a benchmark specifically designed to measure \textsc{Fixturize}’s performance in both classification and fixture-dependent test generation. FixtureEval is first structured into two distinct subsets: $\text{FixtureEval}_\text{G}$, a Python subset derived from real-world open-source GitHub repositories to reflect practical complexity, and $\text{FixtureEval}_\text{L}$, another Python subset manually constructed to prevent data contamination risks associated with LLM pre-training. 
To ensure a comprehensive evaluation, both $\text{FixtureEval}_\text{G}$ and $\text{FixtureEval}_\text{L}$, select programs spanning a wide range of application domains, including algorithm implementation, data science workflows, multimedia processing, natural language processing (NLP), and web scraping.  
Detailed categories, descriptions, and sample counts are summarized in Table \ref{combined-table}. 
To explore the cross-PL capability of \textsc{Fixturize}, we extend FixtureEval with $\text{FixtureEval}_\text{J}$, a Java subset derived from an existing test generation benchmark \cite{Chattester}, which is also composed of real-world code samples. We do not curate code samples from the wild here because test generation benchmarks exclusively for Java are pervasive in the literature and of high quality.
To make readers learn more about FixtureEval, we provide specific statistics in Table \ref{tab:dataset-metrics} to depict \#Samples (the total number of samples), \#Fixture-Dep./-Indep. (the total number of fixture-dependent/Independent samples), \#LOC (the number of lines of code), CC (cyclomatic complexity), \#Tokens (token count per sample), \#Branches (branch count per sample), Depth (AST nesting depth per sample), and \#Imports (imported dependency count per sample).


\begin{table}[ht]
\small
\centering
\caption{Categories, Descriptions, and Sample Counts in FixtureEval (sorted by total count)}
\resizebox{\linewidth}{!}{
\begin{tabular}{l l c c c}
\toprule
\textbf{Category} & \textbf{Description} & \textbf{$\text{FixtureEval}_\text{G}$} & \textbf{$\text{FixtureEval}_\text{L}$} & \textbf{$\text{FixtureEval}_\text{J}$} \\
\midrule
Utilities & General helper utilities, concurrency tools, and program scaffolding. & 8 & 44 & 84\\
Algorithms & Symbolic mathematics, geometric transformation and general algorithmic logic. & 23 & 20 & 4\\
Data Science & Model training, feature processing, and data pipeline construction. & 32 & 10 & 0 \\
Natural Language Processing & Text segmentation, linguistic analysis, and fuzzy matching. & 5 & 26 & 1\\
Multimedia Processing & Programmatic image, audio, and video generation or manipulation. & 35 & 9 & 2\\
Web Scraping & Web data extraction, page parsing, and automated browsing workflows. & 26 & 6 & 1\\
Network Services & Network communication, protocol handling, and service integrations. & 10 & 19 & 71 \\
Data Formats Processing & Reading, converting, and validating structured file formats. & 8 & 14 & 9\\
LLM Training and Serving & Large-model training, inference optimization, and agent execution frameworks. & 24 & 0 & 0 \\
Financial Analysis & Market data analysis, quantitative metrics, and trading strategy logic. & 14 & 0 & 0 \\
File System Operations & Local file handling, compression, traversal, and path management. & 6 & 26 & 9\\
Database & Structured/NoSQL data access, migration, and schema management. & 2 & 18 & 18\\
DevOps & Deployment automation, system operations, and environment management. & 6 & 4 & 0 \\
Security Services & Encryption, signing, and secure access control. & 1 & 4 & 1\\
\bottomrule
\end{tabular}
}
\label{combined-table}
\end{table}

\begin{table}[ht]
\small
\centering
\caption{Summary Statistics of FixtureEval}
\resizebox{\linewidth}{!}{
\begin{tabular}{l c c c c c c c c c}
\toprule
\textbf{Dataset} & \textbf{\#Samples} & \textbf{\#Fixture-Dep.} & \textbf{\#Fixture-Indep.} & 
\textbf{\#LOC} & \textbf{CC} & \textbf{\#Tokens} & \textbf{\#Branches}& \textbf{Depths} & \textbf{\#Imports} \\
\midrule
$\text{FixtureEval}_\text{G}$ 
& $200$ & $100$ & $100$
& $17.61 \pm 11.68$ 
& $4.09 \pm 4.56$
& $211.15 \pm 177.33$
& $2.50 \pm 2.95$
& $9.95 \pm 2.41$
& $2.33 \pm 1.28$ \\
$\text{FixtureEval}_\text{L}$ 
& $200$ & $100$ & $100$
& $18.98 \pm 13.81$
& $3.57 \pm 2.57$
& $139.38 \pm 104.60$
& $2.40 \pm 2.30$
& $8.38 \pm 1.82$
& $1.98 \pm 1.18$ \\
$\text{FixtureEval}_\text{J}$ 
& $200$ & $100$ & $100$
& $21.56 \pm 13.15$
& $4.32 \pm 2.78$
& $205.88 \pm 142.11$
& $2.73 \pm 2.40$
& $1.27 \pm 0.73$
& $3.71 \pm 3.52$ \\
\bottomrule
\multicolumn{10}{l}{\textit{Note:} \#LOC, CC, \#Tokens, \#Branches, Depths and \#Imports are reported as Mean ± Standard Deviation.}
\end{tabular}
}
\label{tab:dataset-metrics}
\end{table}

The construction of FixtureEval is carried out by a team of seven researchers, each with an average of over four years of programming experience and deep expertise in Python and Java, as well as their corresponding test case design. Team roles are clearly defined according to the specific tasks: two primary implementers develop the core program logic, responsible for constructing $\text{FixtureEval}_\text{L}$; two researchers specialize in sourcing and filtering real-world projects from GitHub, responsible for collecting $\text{FixtureEval}_\text{G}$; another two researchers are assigned to select and adapt function-level examples from the dataset used in the empirical study of \textsc{ChatTester} \cite{Chattester}, thereby constructing $\text{FixtureEval}_\text{J}$. Besides, one independent reviewer oversees the verification and quality control for all outputs. 
We elaborate on the detailed construction procedure below.

\begin{table}[ht]
\small
\centering
\caption{GitHub Repository Statistics with Categories and Sample Counts}
\resizebox{\linewidth}{!}{
\begin{tabular}{c c c c c c c}
\toprule
Repository Name & Owner & Category & Stars & Forks & Contributors & Samples \\
\midrule
Python & TheAlgorithms & Algorithms & 210463 & 48574 & 453 & 6 \\
stable-diffusion-webui & AUTOMATIC1111 & Multimedia Processing & 157172 & 29169 & 428 & 32 \\
langflow & langflow-ai & LLM Training and Serving & 130560 & 7769 & 287 & 4 \\
DeepSeek-V3 & deepseek-ai & LLM Training and Serving & 99602 & 16277 & 23 & 1 \\
pytorch & pytorch & Data Science & 93809 & 25512 & 321 & 4 \\
manim & 3b1b & Multimedia Processing & 81093 & 6890 & 164 & 21 \\
screenshot-to-code & abi & Data Formats Processing & 70955 & 8793 & 25 & 1 \\
ansible & ansible & DevOps & 66696 & 24106 & 374 & 2 \\
OpenHands & All-Hands-AI & LLM Training and Serving & 64087 & 7750 & 384 & 1 \\
crawl4ai & unclecode & Web Scraping & 54394 & 5421 & 46 & 31 \\
ChatGPT & lencx & LLM Training and Serving & 54158 & 6174 & 29 & 1 \\
ultralytics & ultralytics & Multimedia Processing & 47071 & 9112 & 305 & 6 \\
diagrams & mingrammer & Multimedia Processing & 41590 & 2682 & 168 & 2 \\
exo & exo-explore & Utilities & 38093 & 2545 & 63 & 3 \\
mindsdb & mindsdb & Data Science & 36358 & 5846 & 381 & 3 \\
spaCy & explosion & Natural Language Processing& 32624 & 4595 & 389 & 1 \\
tinygrad & tinygrad & Data Science & 30244 & 3653 & 414 & 16 \\
redash & getredash & Database & 27854 & 4523 & 389 & 2 \\
PythonRobotics & AtsushiSakai & Algorithms & 26008 & 6892 & 131 & 3 \\
audiocraft & facebookresearch & Multimedia Processing & 22533 & 2455 & 35 & 1 \\
surya & datalab-to & Utilities & 18683 & 1268 & 17 & 1 \\
sqlmodel & fastapi & Database & 16907 & 770 & 94 & 1 \\
hummingbot & hummingbot & Financial Analysis & 14695 & 4001 & 216 & 1 \\
verl & volcengine & LLM Training and Serving & 14153 & 2523 & 354 & 18 \\
claude-code-templates & davila7 & LLM Training and Serving & 13546 & 1182 & 28 & 8 \\
stock & myhhub & Financial Analysis & 10351 & 2057 & 4 & 1 \\
QUANTAXIS & yutiansut & Financial Analysis & 9304 & 3156 & 61 & 7 \\
BlenderGIS & domlysz & Multimedia Processing & 8524 & 1437 & 13 & 2 \\
FATE & FederatedAI & Data Science & 5979 & 1564 & 85 & 2 \\
easyquotation & shidenggui & Financial Analysis & 4810 & 1470 & 17 & 1 \\
Sana & NVlabs & Multimedia Processing & 4554 & 299 & 19 & 1 \\
tqsdk-python & shinnytech & Financial Analysis & 4245 & 711 & 17 & 1 \\
3DDFA & cleardusk & Multimedia Processing & 3671 & 651 & 8 & 2 \\
linkedin\_scraper & joeyism & Web Scraping & 3548 & 829 & 31 & 1 \\
easyquant & shidenggui & Financial Analysis & 3305 & 1241 & 9 & 1 \\
pybroker & edtechre & Financial Analysis & 2817 & 362 & 4 & 2 \\
translationCSAPP & EugeneLiu & Utilities & 2749 & 283 & 44 & 1 \\
LightX2V & ModelTC & Multimedia Processing & 1545 & 105 & 34 & 6 \\
Scweet & Altimis & Web Scraping & 1222 & 245 & 10 & 1 \\
finhack & FinHackCN & Financial Analysis & 810 & 182 & 1 & 1 \\
\bottomrule
\end{tabular}
}
\vspace{0.5em}
\label{tab:repo_stats_distributed}
\end{table}

\subsection{$\text{FixtureEval}_\text{G}$ Collection}
Our data collection for $\text{FixtureEval}_\text{G}$ begins with sourcing focal methods from a curated set of 40 distinct GitHub repositories, as detailed in Table \ref{tab:repo_stats_distributed}. The selection of these projects adheres to the following criteria: (1) we select Python projects from GitHub Trending, prioritizing those with high star counts to ensure popularity and representativeness; (2) the selected projects span multiple domains, ranging from core infrastructure (e.g., Machine Learning, DevOps) to specialized domains (e.g., Algorithmic Trading, Generative AI), and cover projects of varying scales and popularity; (3) we focus on projects that are actively maintained and consistently updated, thereby ensuring higher code quality; and (4) we exclude functions with existing test cases to prevent potential data leakage. This approach ensures the practical value and reliability of the source code. The data collection operates strictly at the function level. Crucially, we ensure that each selected function remains independent.

Based on these rigorous selection principles, we proceed to a meticulous annotation phase. Each qualified function is manually inspected and annotated by expert evaluators to determine its fixture dependency status. To ensure a balanced and focused evaluation set, we continuously monitor the total count of samples in each category (i.e., fixture-dependent and -independent). Data collection concludes once we have secured a sufficient and equal number of high-quality samples for both categories. 
Ultimately, this process yields $\text{FixtureEval}_\text{G}$, a curated benchmark comprising 200 Python functions with 100 fixture-dependent functions and 100 fixture-independent ones. This balanced composition is critical for unbiased evaluation, preventing LLMs' performance from being skewed by class imbalance. 

\subsection{$\text{FixtureEval}_\text{L}$ Construction}
Data contamination poses a significant threat to the validity of LLM evaluations, particularly when using open-source code that may have been included in the LLMs' vast pre-training corpora. Since some of the commercial LLMs studied in this work (see Section \ref{LLMs selection}) are closed-source, it is impossible to verify whether the public functions in $\text{FixtureEval}_\text{G}$ have been memorized, which could unfairly interfere with performance metrics. To rigorously assess generalization capability beyond memorization, we construct an additional evaluation dataset, namely $\text{FixtureEval}_\text{L}$. This subset consists entirely of novel, manually implemented functions that did not exist online before this study, thereby eliminating the possibility of data leakage. We also formulate a clear set of construction principles and strictly adhere to them throughout this process below. 

\textbf{Functional Abstraction and Encapsulation.} Following $\text{FixtureEval}_\text{G}$, we abstract representative functionalities from realistic software scenarios into self-contained functions with clean interfaces. These abstractions draw inspiration from the diverse behaviors and API patterns commonly seen in real-world applications, ensuring coverage of various function types and the preparatory steps they typically require. Although simplified, these functions still depend on fixtures to manage nontrivial runtime states, allowing us to systematically evaluate \textsc{Fixturize}’s ability to generate tests for fixture-dependent functions and to assess whether LLMs can infer implicit dependencies and environmental requirements. 

\textbf{Implicit Environmental Dependencies.} The benchmark methods include carefully designed logic to validate the necessity of comprehensive test preparation. Beyond algorithmic complexity, these methods incorporate complex state management and external interactions that cannot be satisfied by simple parameter passing. Furthermore, the methods embed implicit dependencies where correct functionality relies heavily on the execution environment. For instance, cloud service interactions require not just API calls but properly configured environment variables and network mocks; archive extraction operations depend on the pre-existence of files with specific formats and permissions; and packaging utilities necessitate valid entry scripts and OS-specific build environments. This design forces LLMs to go beyond translating code logic into assertions; they must infer and synthesize the complete execution context required to make the function runnable, which is essentially the invisible fixture code.

\textbf{Coding Practices and Standards.} All original programs strictly adhere to the standard coding conventions and best practices of their respective PLs. This includes clear naming conventions, consistent code formatting, and logical file organization to ensure readability and alignment with real-world development practices. 

Based on these construction principles, we develop the $\text{FixtureEval}_\text{L}$ dataset, which maintains the same scale as $\text{FixtureEval}_\text{G}$ with 100 fixture-dependent and 100 fixture-independent instances. This carefully constructed dataset serves to evaluate \textsc{Fixturize}’s performance in identifying fixture-dependent functions and improving test case generation for such functions under a leakage-free setting. 

\subsection{$\text{FixtureEval}_\text{J}$ Construction} 
A crucial assessment for any new approach is whether its benefits generalize beyond a specific language. To investigate this, we extend FixtureEval with $\text{FixtureEval}_{\text{J}}$, a new subset for evaluating performance in a statically-typed PL, i.e., Java. We curate this benchmark by selecting and adapting function-level examples from four projects in the dataset used in the empirical study of C\textsc{hat}T\textsc{ester} \cite{Chattester}: hutool \cite{hutool}, ormlite-core \cite{ormlite-core}, Flickr4Java \cite{Flickr4Java} and gwt-commons-lang3 \cite{gwt-commons-lang3}. To ensure a direct parallel with our Python evaluations, this dataset maintains the same scale and balanced structure, comprising 100 fixture-dependent and 100 fixture-independent functions. 
The distribution of sample categories is listed in Table \ref{combined-table} and its specific statistics are shown in Table \ref{tab:dataset-metrics}. 
For implementation details, we broadly follow the experimental setup of \textsc{ChatTester}, with \textit{JDK version}=\textit{OpenJDK} 23.0.2 and using Maven 3 \cite{maven} to manage third-party Java dependencies. To ensure consistency across all generated tests, we standardize all tests on JUnit 5.

\section{Evaluation Setup}
This section presents the associated setup for experimentation, such as research questions, datasets, baselines, metrics, LLM selection, and implementation details.
\subsection{Research Questions}

We propose the following research questions (RQs) to comprehensively evaluate \textsc{Fixturize}.

\textbf{RQ1 (Overall Performance): How effective is \textsc{Fixturize} in generating tests on FixtureEval compared to direct LLM prompting? } 
FixtureEval simulates real-world testing scenarios comprising both fixture-dependent and -independent focal methods, on which we can effectively evaluate the performance of \textsc{Fixturize} and make comparisons with baselines. Although \textsc{Fixturize} is capable of combining with existing testing tools, we here only use direct LLM prompting as the baseline to simplify the experimental setting. The performance discussion when integrating with existing testing tools is shown in RQ4.

\textbf{RQ2 (Ablation Study): What is the contribution of each major component to the performance of \textsc{Fixturize}?}
\textsc{Fixturize} consists of multiple components, with each playing a distinct role in handling fixture dependence classification, executable invocation generation, and test generation. Evaluating the contribution of these components is necessary to understand how each part influences the overall framework. 

\textbf{RQ3 (Classification Performance): How accurately does \textsc{Fixturize} discriminate fixture-dependent from fixture-independent functions? }
Accurately distinguishing fixture-dependent from fixture-independent functions is crucial for enabling targeted test generation. It enables \textsc{Fixturize} to focus on fixture-dependent unit test generation and improves overall efficiency.

\textbf{RQ4 (Integration with Existing Approaches): To what extent does \textsc{Fixturize} improve the performance of existing SOTA test generators on FixtureEval? }
\textsc{Fixturize} concentrates on fixture-dependent test generation and incorporates a classification component to route functions appropriately; consequently, it is capable of integrating existing testing tools to handle fixture-independent tests. However, how much can \textsc{Fixturize} improve these existing tools in practice? We hope to offer an answer in this RQ.


\textbf{RQ5 (Cross-Language Generalization): What is the generalization capability of \textsc{Fixturize} across different PLs? }
To understand how well \textsc{Fixturize} performs beyond Python, we evaluate its ability to generalize fixture-dependent test generation across different PLs, reflecting its broader applicability in real-world software development.

\subsection{Datasets}
The experimental evaluation across RQ1 to RQ4 is conducted primarily on $\text{FixtureEval}_{\text{G}}$ and $\text{FixtureEval}_{\text{L}}$, two Python assessment benchmarks introduced in Section \ref{FixtureEval}. As such, we aim to collectively provide a robust assessment by combining real-world challenges with a rigorous safeguard against data leakage. 
To address RQ5 concerning the generalization capability of \textsc{Fixturize} across PLs, we assess it on $\text{FixtureEval}_{\text{J}}$, the Java assessment benchmark.
To enable a focused analysis of \textsc{Fixturize}'s performance on fixture-dependent samples, we isolate the corresponding data for detailed discussion within their respective RQs.

\subsection{Baselines}
\label{baselines}
For RQ1, as mentioned before, although \textsc{Fixturize} is capable of combining with existing testing tools, we employ direct LLM prompting here as a baseline to simplify the experimental setup. Specifically, the direct prompting template is shown in Figure \ref{prompt:direct}, representing the LLM's inherent test generation capability on FixtureEval in our experimental context without any enhancements.
To evaluate the performance of identifying fixture dependence in RQ3, we introduce direct LLM classification to benchmark against IBC. As depicted in Figure \ref{prompt:abstract reasoning}, we instruct the LLM to perform a direct binary classification, asking it to output solely ``yes'' or ``no'' regarding whether the given focal method can be correctly invoked by parameter passing without any previous setup.
For RQ4, which examines \textsc{Fixturize}'s ability to enhance existing SOTA test generators, we select C\textsc{over}U\textsc{p} as a representative LLM-based test generation technique and \textsc{Pynguin} as a traditional search-based approach. By selecting these approaches, we can evaluate how \textsc{Fixturize} enhances diverse testing tools, especially for fixture-dependent functions. 

\begin{figure}[htbp]
\centering

\begin{minipage}{\textwidth}
\begin{promptbox}[title=Prompt : Direct LLM prompting. ]
\textcolor{instr}{Based on the following code, please use 'unittest' to generate a Python test suite that includes 5 test cases. Import each focal method from the specified file using the syntax: `from \{base\_name\} import <func1>, <func2>···'.\\
\textcolor{instr}{code: \{code\}}}
\end{promptbox}
\end{minipage}

\caption{Direct LLM Prompting Template for Test Generation }
    \label{prompt:direct}
\end{figure}
\begin{figure}[htbp]
\centering

\begin{minipage}{\textwidth}
\begin{promptbox}[title=Prompt : Guiding the LLM to perform direct classification. ]
\textcolor{instr}{Determine whether the function can be correctly invoked solely by parameter passing without any previous setup.}\\
\textcolor{code}{code: \{code\}}\\
\textcolor{output}{The output must be only "yes" or "no" without any other words.}
\end{promptbox}
\end{minipage}

\caption{Direct LLM Classification Prompt in RQ3 }
    \label{prompt:abstract reasoning}
\end{figure}

\subsection{Metrics}
\paragraph{(1) Classification.} To evaluate the classification performance of \textsc{Fixturize} and the direct LLM prompting approach, we employ Precision, Accuracy, Recall, and F1-score as standard metrics.

\textbf{Precision} measures the fraction of true fixture-dependent functions among all cases classified as such. It is calculated as: \(\frac{TP}{TP + FP}\), where TP denotes the number of fixture-dependent functions correctly classified, and FP represents the number of fixture-independent functions mistakenly classified as fixture-dependent.

\textbf{Accuracy} evaluates the performance that how many functions can be correctly classified. It is calculated as: \(\frac{TP+TN}{TP + FP+TN+FN}\), where TN represents the number of fixture-independent functions correctly classified and FN represents the number of fixture-dependent functions that were mistakenly classified as fixture-independent.

\textbf{Recall} measures how many fixture-dependent functions can be correctly classified. It is calculated as: \(\frac{TP}{TP + FN}\).

\textbf{F1-score} provides the harmonic mean of Precision and Recall, balancing both concerns. It is calculated as: \(2 \times\frac{Precision \times Recall}{Precision + Recall}\).

\paragraph{(2) Unit Test Generation.} The evaluation of generated test cases follows prior research \cite{RetrievalAugmentedTG} by combining static analysis with dynamic execution, using the following metrics outlined.

\textbf{Parse Rate} (abbr. PR\%) measures the syntactic correctness of generated unit tests .It is defined as below.
\begin{equation}
    PR = \frac{\sum^{N_{x}}_{i=1}pr(\hat{s_{i}})}{N_{x}}, \ \textbf{where} \ pr(\hat{s_{i}}) = \left\{
\begin{matrix}
1 & AST(\hat{s_{i}}) \to success\\
0 & AST(\hat{s_{i}})  \to failure\\
\end{matrix}\right. 
\end{equation}
where $N_{x}$ denotes the total number of test suites, and $\hat{s_{i}}$ denotes the $i$-th generated test suite via a certain LLM. \textit{AST($\cdot$)} refers to the process of using Abstract Syntax Tree parsing to analyze a generated test suite, with an outcome of either \textit{success} or \textit{fail}. It is assessed at the test suite level since LLMs are prompted to generate directly runnable and standalone test suites. 

\textbf{EXecution rate} (abbr. EX\%) reflects the correctness of the test case execution. It is defined as below.
\begin{equation}
    EX = \frac{\sum^{N_{x}}_{i=1}\sum^{T_{i}}_{j=1}ex(\hat{c_{i,j}})}{\sum^{N_{x}}_{i=1} {T_{i}}}, \ \textbf{where} \ ex(\hat{c_{i,j}}) = \left\{
\begin{matrix}
1 \ & \hat{c_{i,j}}\  executes\ without\ raising\ errors   \\
0 \ & \hat{c_{i,j}}\  raises\ execution\ errors   \\
\end{matrix}\right. 
\end{equation}
where $T_{i}$ denotes the total number of test cases in the $i$-th test suite, and $\hat{c_{i,j}}$ denotes the $j$-th test case in the $i$-th test suite.

\textbf{Case PaSs rate} (abbr. CasePS\%) shows the proportion of test cases that pass successfully. It is defined as below.
\begin{equation}
    CasePS = \frac{\sum^{N_{x}}_{i=1}\sum^{T_{i}}_{j=1}case(\hat{c_{i,j}})}{\sum^{N_{x}}_{i=1} {T_{i}}}, \ \textbf{where} \ case(\hat{c_{i,j}}) = \left\{
\begin{matrix}
1 & Exec_{j}(s_i)=Exec_{j}(\hat{s_i})\\
0 & Exec_{j}(s_i)\neq Exec_{j}(\hat{s_i}) \\
\end{matrix}\right. 
\end{equation}
where $s_i$ denotes the ground truth of the  $i$-th test suite, $Exec_{j}(\cdot)$ denotes the execution result  of a test suite ($s_i$ or $\hat{s_i}$) on the $j$-th test case.

\textbf{Suite PaSs rate} (abbr. SuitePS\%) measures the proportion of fully passing test suites, which is also assessed at the test suite level similar to Parse Rate. It is defined as below.
\begin{equation}
    SuitePS = \frac{\sum^{N_{x}}_{i=1}suite(\hat{s_{i}})}{N_{x}}, \ \textbf{where} \ suite(\hat{s_{i}}) = \left\{
\begin{matrix}
1 &  \sum^{T_{i}}_{j=1}case(\hat{c_{i,j}}) = T_i\\
0 &  \sum^{T_{i}}_{j=1}case(\hat{c_{i,j}}) \neq T_i\\
\end{matrix}\right. 
\end{equation}
It provides a stringent assessment of test suite quality by requiring every test case in a suite to pass, thereby indicating the functional coherence and stability of the tested suite.

\textbf{Coverage Metrics.} To facilitate comparison with SOTA approaches in RQ4, we additionally employ two standard test metrics:
\begin{itemize}
\item \textbf{Line Coverage} (abbr. LCov\%) measures the percentage of executable lines covered by the test suite.
\item \textbf{Branch Coverage} (abbr. BCov\%) evaluates the coverage of various code branches, and gages the effectiveness of the test suite in exploring different execution paths. 
\end{itemize}

\subsection{LLM Selection}
\label{LLMs selection}
To assess how \textsc{Fixturize} performs across different LLMs, we employ multiple SOTA LLMs as backbones, including DeepSeek-V3 (hereafter DeepSeek), GPT-4o, Claude-3-5-sonnet-20241022 (hereafter Claude), and Qwen-max-2025-01-25 (hereafter Qwen). DeepSeek and GPT‑4o are among the most advanced and accessible LLMs, often serving as benchmarks in comparative studies for code and reasoning tasks \cite{sadik2025benchmarkingllmcodesmells,10549472}. Claude has demonstrated strong performance in software engineering and general autonomy tasks, achieving results comparable to humans in long-context reasoning and coding workflows \cite{METR2024_Claude3.5_Sonnet_Report}. Qwen, built with a Mixture-of-Experts architecture, often matches or surpasses other top LLMs in general reasoning and coding tasks \cite{QwenMax2025}.

\subsection{Implementation Details}
The evaluation is conducted on Python 3.12.0 for both $\text{FixtureEval}_{\text{G}}$ and $\text{FixtureEval}_{\text{L}}$. We use Docker to evaluate C\textsc{over}U\textsc{p} \cite{coverup_github}, where the environment is based on Miniconda3 with Python 3.12.0. For \textsc{Pynguin}, we use Python 3.10.18 packaged by Anaconda, as \textsc{Pynguin} requires older versions of Python. 

We use the four LLMs mentioned above as the backbone for our experiments, all accessed through their official API interfaces \cite{deepseek,openai,qwen-aliyun,claude}. We use the following hyperparameters: \textit{temperature}=0 and \textit{n}=1 for both classification and unit test generation, along with all other hyperparameters kept by default. This means that during inference, we only collect the first candidate generated by LLMs, thereby effectively minimizing the randomness of the LLM and ensuring the reproducibility of the experiments. Nonetheless, despite this, the randomness persists to some extent. Thus, we assessed stability by conducting small-scale repeated experiments multiple times following previous studies \cite{Third-Party,LLMorpheus}
Specifically, we carry out repeated experiments three times with each LLM to mandate them to generate unit tests on $\text{FixtureEval}_{\text{L}}$ with direct prompting.
The result reveals that all metrics in all LLMs except CasePS in Qwen exhibit a variance of less than 3\%. As for the CasePS metric in Qwen, the variance is around 3.08\%. Therefore, the overall low variance observed in this analysis indicates that the inherent randomness has a limited impact on the replication of our study. 


\definecolor{light cyan}{HTML}{E1FFFF}
\begin{table}[htbp]
\centering
\caption{Overall Performance Evaluation of \textsc{Fixturize} on FixtureEval}
\resizebox{\linewidth}{!}{
\begin{tabular}{llcccccccc}
\toprule
\multirow{2}{*}{\textbf{LLM}}& \multirow{2}{*}{\textbf{Approach}}& \multicolumn{4}{c}{\textbf{$\text{FixtureEval}_{\text{G}}$}} & \multicolumn{4}{c}{\textbf{$\text{FixtureEval}_{\text{L}}$}} \\
\cmidrule(lr){3-6} \cmidrule(lr){7-10}
& & \textbf{PR} & \textbf{EX} & \textbf{CasePS}& \textbf{SuitePS} & \textbf{PR} & \textbf{EX} & \textbf{CasePS}& \textbf{SuitePS} \\
\midrule
\multirow{3}{*}{DeepSeek} 
& Direct & 98.00\% (97.09\%)& 92.71\% (88.68\%)& 58.83\% (47.49\%)& 32.00\% (24.27\%)& 99.50\% (\textbf{100.00\%}) & 95.78\% (92.76\%) & 75.07\% (66.93\%) & 45.50\% (35.00\%) \\
& \textsc{Fixturize} & 99.00\% (99.03\%)& 97.09\% (97.15\%)& 70.46\% (69.96\%)& 45.50\% (50.48\%)& 99.50\% (\textbf{100.00\%}) & 97.78\% (97.81\%) & 85.69\% (89.24\%) & 65.00\% (75.00\%) \\[-1pt] 
& Imprv. & \colorbox{light cyan}{1.02\% (2.00\%)}& \colorbox{light cyan}{4.72\% (9.55\%)}& 19.77\% (47.30\%)& 42.19\% (108.00\%)& 0.00\% (0.00\%)& 2.08\% \colorbox{light cyan}{(5.44\%)} & 14.14\% (33.34\%) & 42.86\% (114.29\%) \\[-1pt] 
\midrule
\multirow{3}{*}{GPT-4o} 
& Direct & \textbf{100.00\%} (\textbf{100.00\%}) & 93.45\% (88.51\%)& 58.69\% (46.89\%)& 32.00\% (23.81\%)& \textbf{100.00\%} (\textbf{100.00\%}) & 98.50\% (96.88\%) & 71.76\% (62.71\%) & 42.00\% (31.00\%) \\
& \textsc{Fixturize} & \textbf{100.00\%} (\textbf{100.00\%}) & 97.52\% (96.23\%)& 69.71\% (67.80\%)& 44.00\% (46.67\%)& \textbf{100.00\%} (\textbf{100.00\%}) & \textbf{99.00\%} (\textbf{97.92\%}) & 81.64\% (83.33\%) & 58.50\% (65.00\%) \\[-1pt] 
& Imprv. & 0.00\% (0.00\%)& 4.36\% (8.72\%)& 18.78\% (44.58\%)& 37.50\% (96.00\%)& 0.00\% (0.00\%)& 0.51\% (1.08\%) & 13.77\% (32.89\%) & 39.29\% (109.68\%) \\[-1pt] 
\midrule
\multirow{3}{*}{Claude} 
& Direct & 98.98\% (98.92\%)& 95.77\% (94.78\%)& 78.07\% (72.65\%)& 55.56\% (51.61\%)
& 95.50\% (96.94\%) & 92.03\% (93.75\%) & 75.10\% (68.49\%) & 53.00\% (41.00\%) \\
& \textsc{Fixturize} & 99.49\% (\textbf{100.00\%})& 97.21\% (97.85\%)& \textbf{82.35\% }(\textbf{81.72\%})& \textbf{65.66\%}(\textbf{73.12\%})& 97.00\% (\textbf{100.00\%}) & 94.53\% (95.97\%) & \textbf{86.17\%} (\textbf{90.73\%}) & \textbf{74.00\%} (\textbf{82.00\%}) \\[-1pt] 
& Imprv. & 0.51\% (1.09\%)& 1.50\% (3.24\%)& 5.48\% (12.48\%)& 18.18\% (41.67\%)
& \colorbox{light cyan}{1.57\% (3.16\%)} & \colorbox{light cyan}{2.71\%} (2.37\%) & 14.74\% (32.48\%) & 39.62\% (100.00\%) \\[-1pt] 
\midrule
\multirow{3}{*}{Qwen} 
& Direct & \textbf{100.00\%} (\textbf{100.00\%}) 
& 96.98\% (95.10\%)& 52.16\% (42.74\%)& 20.50\% (12.62\%)
& \textbf{100.00\%} (\textbf{100.00\%}) & 97.50\% (93.81\%) & 66.40\% (58.35\%) & 34.50\% (23.00\%) \\
& \textsc{Fixturize} & \textbf{100.00\%} (\textbf{100.00\%}) & \textbf{98.50\%} (\textbf{98.05\%})& 64.83\% (67.45\%)& 35.50\% (41.75\%)
& \textbf{100.00\%} (\textbf{100.00\%}) & 98.01\% (94.90\%) & 77.51\% (81.22\%) & 52.50\% (60.00\%) \\[-1pt] 
& Imprv. & 0.00\% (0.00\%)& 1.56\% (3.10\%)& \colorbox{light cyan} {24.29\% (57.79\%)}& \colorbox{light cyan} {73.17\% (230.78\%)}& 0.00\% (0.00\%)& 0.52\% (1.16\%) & \colorbox{light cyan}{16.73\% (39.20\%)} & \colorbox{light cyan}{52.17\% (160.87\%)} \\[-1pt] 
\midrule
\multirow{3}{*}{Average} 
& Direct & 99.25\% (99.00\%)& 94.73\% (91.77\%)& 61.94\% (52.45\%)& 35.01\% (28.08\%)
& 98.75\% (99.23\%) & 95.95\% (94.30\%) & 72.08\% (64.12\%) & 43.75\% (32.50\%) \\
& \textsc{Fixturize} & 99.62\% (99.76\%)& 97.58\% (97.32\%)& 71.84\% (71.73\%)& 47.66\% (53.00\%)
& 99.13\% (\textbf{100.00\%}) & 97.33\% (96.65\%) & 82.75\% (86.13\%) & 62.50\% (70.50\%) \\
& Imprv. & 0.38\% (0.76\%)& 3.01\% (6.05\%)& 15.98\% (36.77\%)& 36.13\% (88.77\%)& 0.38\% (0.77\%) & 1.44\% (2.49\%) & 14.80\% (34.33\%) & 42.86\% (116.92\%) \\
\bottomrule
\multicolumn{10}{p{26cm}}{\textit{Note:} Values in parentheses exclusively represent performance on fixture-dependent samples. Direct refers to the direct LLM prompting approach, and Imprv. refers to the improvement rate. Besides, we use \colorbox{light cyan}{light cyan} background in table cells to denote the highest improvement and \textbf{bold} values to highlight the highest rate in terms of PR/EX/CasePS/SuitePS.}
\end{tabular}
}
\label{Table 2}
\end{table}

\section{Evaluation Results}
This section discusses the experimental results of \textsc{Fixturize}.
\subsection{RQ1: Overall Performance}
\label{RQ1}
\textbf{\textsc{Fixturize} vs. Direct LLM prompting.} 
Table \ref{Table 2} demonstrates the performance comparison between \textsc{Fixturize} and direct LLM prompting on $\text{FixtureEval}_{\text{G}}$ and $\text{FixtureEval}_{\text{L}}$. In particular, their performance exclusively on fixture-dependent samples is listed in associated parentheses.
As can be seen, \textsc{Fixturize} substantially raises the overall test correctness across all evaluated LLMs compared with the direct LLM prompting approach. To be specific, \textsc{Fixturize} on average improves CasePS from 61.94\% to 71.84\%, and SuitePS from 35.01\% to 47.66\% on $\text{FixtureEval}_{\text{G}}$.

As for $\text{FixtureEval}_{\text{L}}$, the improvement of SuitePS and CasePS remains significant, with CasePS increasing from 72.08\% to 82.75\% and SuitePS from 43.75\% to 62.50\%. Critically, this improvement is even more evident when focusing exclusively on fixture-dependent functions (values in parentheses). Specifically, \textsc{Fixturize} on average increases CasePS from 52.45\% to 71.73\% and SuitePS from 28.08\% to 53.00\% on $\text{FixtureEval}_{\text{G}}$. Similarly, on $\text{FixtureEval}_{\text{L}}$, the improvement rates for fixture-dependent samples rise to 34.33\% for CasePS and 116.92\% for SuitePS, confirming that \textsc{Fixturize}'s gains are driven by its efficacy in resolving complex fixture dependencies. This phenomenon also aligns with our failure analysis, which reveals that the proportion of fixture-related failures drops significantly from 31.78\% in direct LLM prompting to 16.96\% with \textsc{Fixturize}.
On metrics evaluating syntactic validity and executability, \textsc{Fixturize} yields marginal improvements over direct LLM prompting. This is expected, as all LLMs already achieve PR over 95\% and EX over 92\% across both benchmarks, indicating that SOTA LLMs are proficient at generating syntactically correct and compilable code, regardless of fixture complexity.
\vspace{-0.3em}
\begin{boxK}
\small \faIcon{pencil-alt} \textbf{Finding 1:} 
\textsc{Fixturize} significantly enhances test correctness (i.e., CasePS and SuitePS), especially when exclusively focusing on fixture-dependent functions.
PR and EX remain consistently high across all settings, demonstrating that it is already easy for SOTA LLMs to generate syntactically correct and executable code.
\end{boxK}
\vspace{-0.3em}

\textbf{Comparison among benchmarks.} 
\textbf{Comparison among benchmarks.}
On average, \textsc{Fixturize} with diverse LLMs demonstrates similar improvement rates across all evaluation metrics and two benchmarks, with an average lifting of 0.38\% on PR, 1.44\%--3.01\% on EX, 14.80\%--15.98\% on CasePS, and 36.13\%--42.86\% on SuitePS. When focusing exclusively on fixture-dependent samples, the average improvement rates become more prominent, with 0.76\%--0.77\%, 2.49\%--6.05\%, 34.33\%--36.77\%, and 88.77\%--116.92\% in terms of each metric in order, showing the consistent effectiveness of \textsc{Fixturize}.
Nevertheless, the absolute scores that \textsc{Fixturize} reach for each benchmark are distinctive. For example, \textsc{Fixturize} obtains a CasePS score of 82.75\% and a SuitePS score of 62.50\% on $\text{FixtureEval}_{\text{L}}$, while it only achieves a CasePS score of 71.84\% and a SuitePS score of 47.66\% on $\text{FixtureEval}_{\text{G}}$. In addition, it is also the case for exclusively fixture-dependent samples, where \textsc{Fixturize} reaches 86.13\% CasePS and 70.50\% SuitePS on $\text{FixtureEval}_{\text{L}}$, but only 71.73\% CasePS and 53.00\% SuitePS on $\text{FixtureEval}_{\text{G}}$, suggesting that the real-world coding problems are more challenging.
This phenomenon is explainable according to our statistics in Table \ref{tab:dataset-metrics}, compared with $\text{FixtureEval}_{\text{L}}$, $\text{FixtureEval}_{\text{G}}$ exhibits higher cyclomatic complexity (4.09 vs. 3.57) and token count (211.15 vs. 139.38), indicating that the latter contains more semantically denser and structurally intricate code samples. Besides, the latter depends on more external modules (2.33 vs. 1.98), leading to higher coupling code structures. Therefore, all of the above factors collectively aggravate the difficulty of generating tests for $\text{FixtureEval}_{\text{G}}$ \cite{Badri2012_EmpiricalTestability,Chowdhury2011_CCC,Alawad2019_ReadabilityComplexity,10.5555/3737916.3742190}.
However, as for PR and EX, they already reach near-ceiling performance, with PR above 97.0\% and EX above 94.5\%, across both benchmarks, indicating that the syntactic validity and executability of generated tests are not substantially affected by dataset complexity. 

\vspace{-0.3em}
\begin{boxK}
\small \faIcon{pencil-alt} \textbf{Finding 2:} 
Generating tests for $\text{FixtureEval}_{\text{G}}$ is more challenging than for $\text{FixtureEval}_{\text{L}}$, as it contains focal methods with higher semantic density and coupling. Besides, the vast majority of generated tests are syntactically valid and executable regardless of dataset complexity.
\end{boxK}
\vspace{-0.3em}

\textbf{Comparison among LLMs.} 
Overall, \textsc{Fixturize} with Claude performs the best.
To be specific, on SuitePS, Claude achieves an average score of 69.83\% with \textsc{Fixturize} across both benchmarks and outperforms DeepSeek (55.25\%), GPT-4o (51.25\%), and Qwen (44.00\%) by 26.39\%, 36.25\%, and 58.70\%, respectively. On CasePS, Claude reaches an average score of 84.26\%, surpassing DeepSeek (78.07\%), GPT-4o (75.67\%), and Qwen (71.17\%) with relative gains of 7.92\%, 11.34\%, and 18.39\%. Claude’s superior performance aligns with prior findings \cite{nascimento2024llm4ds,10.1145/3728940,wei2025invbenchllm} and is also demonstrated in the fixture-dependent scenarios. 
In addition, we found that Qwen obtains the lowest scores with direct prompting, but achieves the largest relative gains with \textsc{Fixturize} by 16.73\%--24.29\% in terms of CasePS, and 52.17\%--73.17\% in terms of SuitePS across both benchmarks. 
On the fixture-dependent subset (parenthesized values), these gains jump to 39.20\%--57.79\% and 160.87\%--230.78\%, respectively, highlighting the effectiveness of the \textsc{Fixturize} in elevating LLMs with relatively poor initial performance. Notably, \textsc{Fixturize} remains beneficial even for high-starting LLMs such as Claude, which achieves 5.48\%--14.74\% improvement in CasePS and 18.18\%--39.62\% in SuitePS across both benchmarks. 
For PR and EX, all LLMs achieve PR above 97\% and EX above 94.5\% across the board, reflecting their strong capability in generating syntactically valid and executable test code. 

\vspace{-0.3em}
\begin{boxK}
\small \faIcon{pencil-alt} \textbf{Finding 3:} 
Claude achieves the highest SuitePS and CasePS scores, while LLMs with poor performance, such as Qwen, obtain the largest relative gains, demonstrating that \textsc{Fixturize} benefits both high- and low-performing LLMs. Moreover, all LLMs perform well on code syntactic validity and executability. 
\end{boxK}
\vspace{-0.3em}

\subsection{RQ2: Ablation Study}
\label{RQ2}
RQ2 evaluates how each major component of \textsc{Fixturize} contributes to the overall performance. We evaluate the framework from two perspectives: (1) the impact of \textsc{Fixturize}'s classification, by comparing it with the random classification; and
(2) the contribution of each generation module, through a stepwise ablation study. 
To achieve an appropriate trade-off between experimental reliability and computational overhead, we select two of the LLMs (i.e., DeepSeek and GPT-4o) under test for this evaluation.

\begin{figure}
    \centering
    \includegraphics[width=1\linewidth]{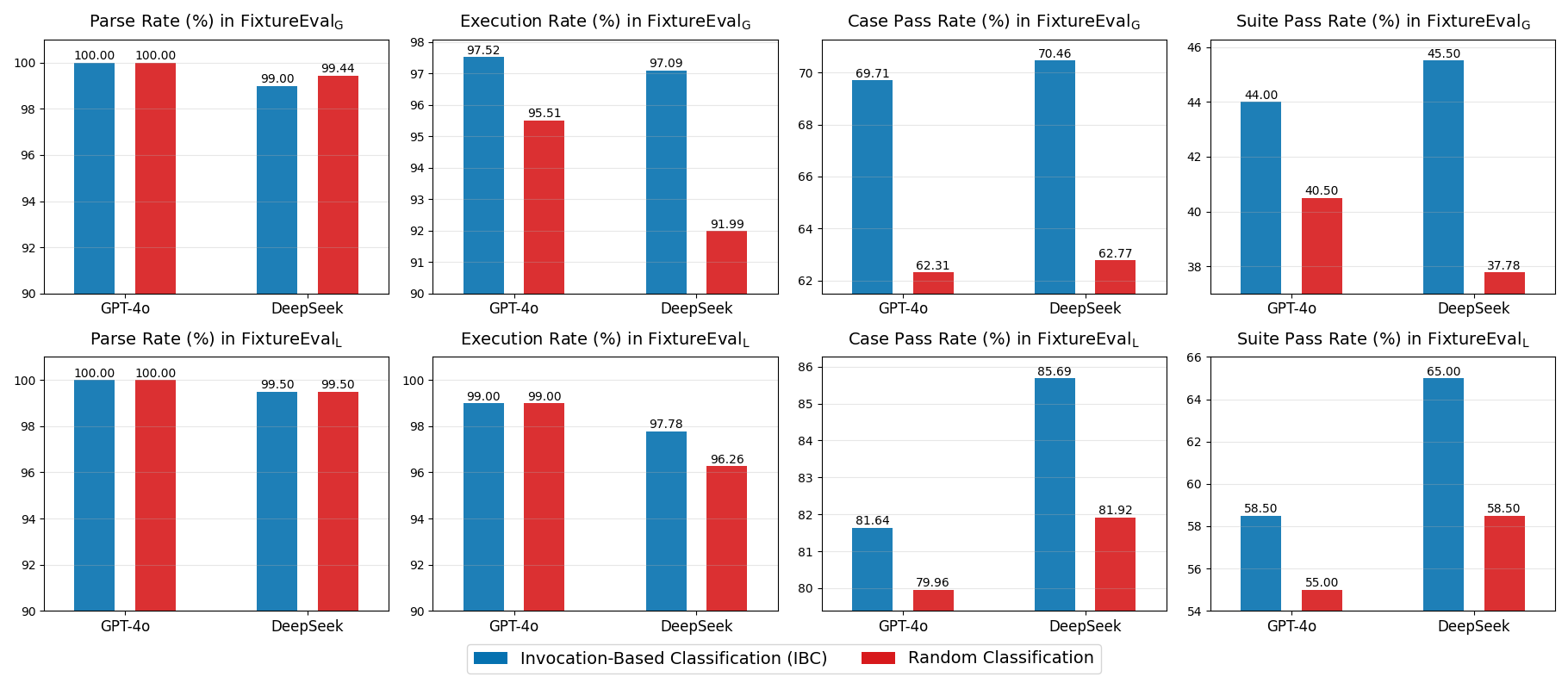}
    \caption{IBC vs. Random Classification on Test Generation Effectiveness }
    \label{fig:RQ2-random}
    
\end{figure}

\textbf{Ablation Study on Classification.} 
We first examine the classification component of \textsc{Fixturize}, which determines whether a focal method should proceed through fixture-dependent test generation or not. To evaluate this section, we compare the IBC of \textsc{Fixturize} against the random classification (seed = 123456) while keeping the test generation process identical across both settings. As shown in Figure \ref{fig:RQ2-random}, IBC consistently matches or outperforms the random classification at nearly all metrics on both DeepSeek and GPT-4o. On $\text{FixtureEval}_{\text{L}}$, DeepSeek shows clear gains, raising CasePS from 81.92\% to 85.69\% and SuitePS from 58.50\% to 65.00\%, while GPT-4o exhibits similar improvements, increasing CasePS from 79.96\% to 81.64\% and SuitePS from 55.00\% to 58.50\%. As for PR and EX, \textsc{Fixturize} with both classification methods has reached saturated performance, making the improvement efficacy of IBC limited, which is consistent with the observation in Section \ref{RQ1}. On the $\text{FixtureEval}_{\text{G}}$, the performance gain brought by classification is consistent with that on $\text{FixtureEval}_{\text{L}}$. For DeepSeek, IBC boosts CasePS from 62.77\% to 70.46\% and SuitePS from 37.78\% to 45.50\% . Similarly, the CasePS of GPT-4o rise from 62.31\% to 69.71\% and SuitePS from 40.50\% to 44.00\%.



 

\vspace{-0.3em}
\begin{boxK}
\small \faIcon{pencil-alt} \textbf{Finding 4:} 
IBC provides clear benefits on both $\text{FixtureEval}_{\text{L}}$ and $\text{FixtureEval}_{\text{G}}$, significantly improving CasePS and SuitePS over the random classification, while PR and EX remain almost saturated. 
\end{boxK}
\vspace{-0.3em}

\begin{figure}
    \centering
    \includegraphics[width=1\linewidth]{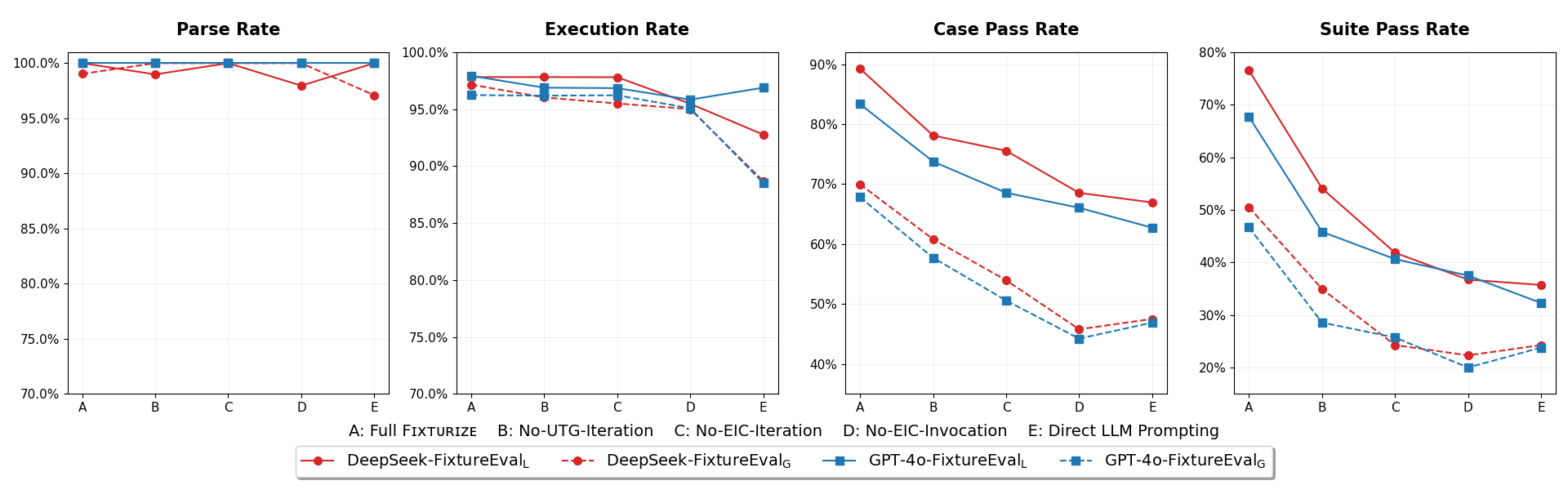}
    \caption{Ablation Study of Test Generation Components }
    \label{fig:RQ2-ablation}
\end{figure}

\textbf{Ablation Study on Test Generation.} 
Considering that the follow-up components of \textsc{Fixturize} only affect the test generation performance of samples classified as fixture-dependent by IBC, we carry out systematic ablation studies on this subset. 
Overall, apart from the IBC module, \textsc{Fixturize} consists of an Executable Invocation Construction (EIC) and a Unit Test Generation (UTG) module, where the former comprises two sub-modules: one is for (1) executable function-call generation, and another is for (2) iterative repair augmented with mocking hints. As for the UTG module, it is also composed of two sub-modules, i.e., (1) UTG with fixture hints and (2) iterative repair with execution feedback.
To achieve a systematic ablation experiment, we define five configurations (from A to E), progressively stripping away components from the full framework back to the baseline:
\begin{itemize}
\item \textbf{Configuration A (Full \textsc{Fixturize}):} The complete \textsc{Fixturize} approach, including all components from IBC, EIC, and UTG.
\item \textbf{Configuration B (No-UTG-Iteration):} Removes the iterative repair loop within the UTG module. The LLM must generate the test suite in a single pass based on invocation examples and fixture hints. 
\item \textbf{Configuration C (No-EIC-Iteration):} Building on Configuration B, this step further removes the iterative repair with error message and mocking hints within the EIC module, making the LLMs only rely on initial function-call generation and fixture hints for test generation.
\item \textbf{Configuration D (No-EIC-Invocation):} Configuration C further removes the invocation examples (i.e., the executable code snippets generated in EIC). The LLM is guided solely by fixture hints.
\item \textbf{Configuration E (Direct LLM Prompting):} Removes all guidance (including fixture hints), reverting to direct LLM prompting.
\end{itemize}
Figure \ref{fig:RQ2-ablation} shows the ablation results of the test generation process on both $\text{FixtureEval}_{\text{L}}$ and $\text{FixtureEval}_{\text{G}}$ benchmarks using DeepSeek and GPT-4o. Overall, the performance gradually degrades as components are removed, confirming that each module makes a positive contribution to the generation quality.
For example, the average reduction on CasePS is 12.88\% when the UTG-Iteration (B) is removed, followed by smaller drops of 8.04\% (C), 9.68\% (D), and 0.23\% (E) in subsequent ablation steps. SuitePS exhibits an even more evident decline with each ablated version on average, falling by 32.30\% (B), 18.96\% (C), 12.00\% (D), and 0.41\% (E), respectively. These results indicate that all components contribute to test quality overall, with UTG-Iteration and EIC-Invocation having the most substantial impact. Meanwhile, PR remains saturated across all ablations, with GPT-4o achieving a score of 100\% on PR across both benchmarks, causing their curves to overlap. EX remains largely neck-to-neck across ablations, showing only gradual declines. This demonstrates that for SOTA LLMs, generating syntactically valid and executable code has become a baseline capability. The true challenge lies in bridging the gap to functional correctness and logic sufficiency.

\vspace{-0.3em}
\begin{boxK}
\small \faIcon{pencil-alt} \textbf{Finding 5:} 
The stepwise ablation reveals that \textsc{Fixturize}’s performance degrades consistently as modules are removed. While PR and EX remain stable, CasePS and SuitePS drop substantially at each step, showing that these components are essential for producing functionally correct tests.
\end{boxK}
\vspace{-0.3em}

\textbf{Anomaly Analysis.} 
When interpreting these results, it is essential to recognize that the components are not entirely independent \cite{altmayer2025coverup}.  For example, we observe that the configuration D (No-EIC-Invocation) often performs comparably to, or even slightly worse than, the configuration E (Direct LLM Prompting) in terms of CasePS and SuitePS, especially on $\text{FixtureEval}_{\text{G}}$. 
We then analyzed the failure cases using configuration D in $\text{FixtureEval}_{\text{G}}$ to explore this anomaly. Among them, we found that 58.01\% of errors result from incorrect fixture setup, where most of them arise from a variety of defects largely introduced by the added fixture hint. This suggests that while the fixture hint is intended to guide test generation, it also increases task complexity, introducing new failure points. Incorporating invocation examples in Configuration C mitigates these issues, enabling the LLM to rely on concrete context rather than speculation, thereby focusing on generating high-quality tests. This reveals that fixture hints are ineffective without concrete invocations, highlighting the essential synergy between these components.


 \vspace{-0.3em}
\begin{boxK}
\small \faIcon{pencil-alt} \textbf{Finding 6:} 
While individual components may introduce specific defects, the synergistic effect among \textsc{Fixturize} modules effectively mitigates these risks. The full \textsc{Fixturize} creates a self-correcting mechanism, ensuring syntactically valid and executable tests where isolated components fail.
\end{boxK}
\vspace{-0.3em}

\subsection{RQ3: Classification Performance }
\label{RQ3}

\begin{figure}
    \centering
    \includegraphics[width=1\linewidth]{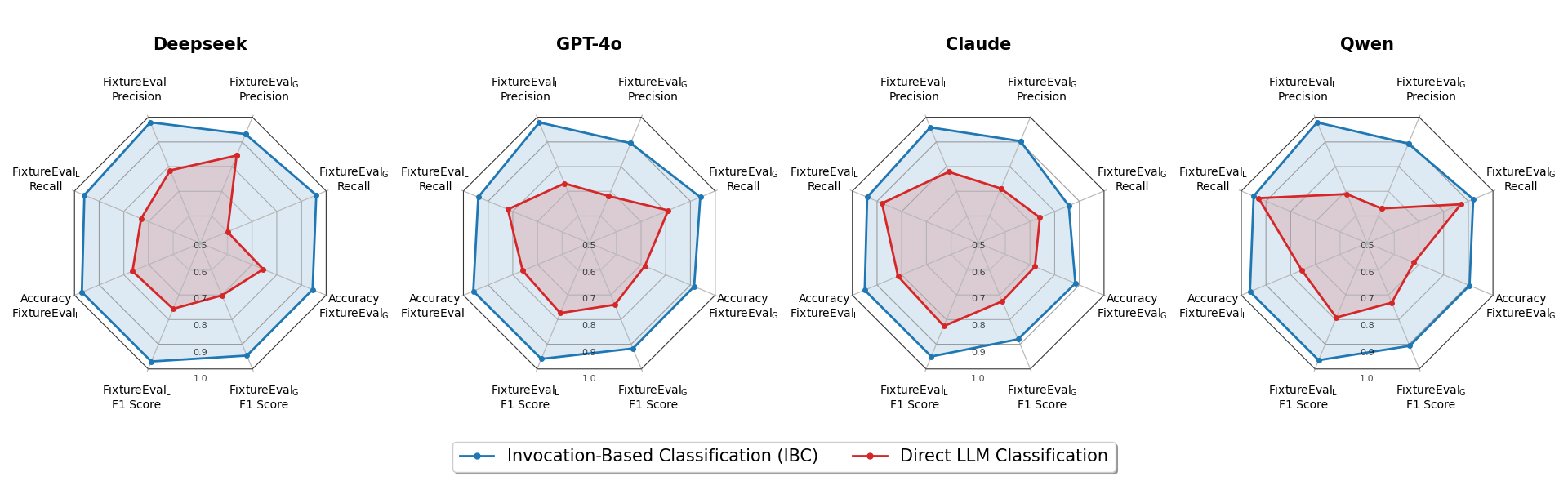}
    \caption{Comparative Performance of Direct LLM Classification and IBC Across LLMs and Datasets}
    \label{fig:radar plot}
\end{figure}
\textbf{IBC vs. Direct LLM Classification.} We evaluate the classification capabilities of IBC against direct LLM classification across $\text{FixtureEval}_{\text{L}}$ and $\text{FixtureEval}_{\text{G}}$. The performance metrics include precision, recall, accuracy, and F1 score, as summarized in Figure \ref{fig:radar plot}, demonstrating that IBC consistently outperforms direct LLM classification across all datasets and LLMs. On $\text{FixtureEval}_{\text{L}}$, IBC outperforms the direct classification on average by 30.58\% in terms of precision, 12.80\% in terms of recall, 24.23\% in terms of accuracy, and 22.08\% in terms of F1 score across all LLMs. As for the more challenging benchmark, $\text{FixtureEval}_{\text{G}}$, IBC still maintains the superiority on all LLMs, where it improves the recall score by 21.76\% and F1 score by 25.91\% on average. Besides, the performance gain remains consistently high in terms of precision and accuracy, obtaining 26.36\% and 27.58\% lifting on average, respectively.  
As can be seen, direct LLM classification requires the LLM to interpret complete code structures, control flows, and semantic dependencies, which is an abstractive reasoning task that current LLMs find hard to handle. In contrast, IBC treats method classification as a simple method invocation task, where LLMs only need to concentrate on the function signature instead of the whole function semantics. In addition, the final classification part depends on the concrete execution result, significantly enhancing the decision reliability and certainty. 



\vspace{-0.3em}
\begin{boxK}
\small \faIcon{pencil-alt} \textbf{Finding 7:} 
By reducing the task to generating a single-line invocation, IBC removes the noise and cognitive burden of analyzing full code, yielding consistently much higher performance compared with direct LLM classification across all LLMs.
\end{boxK}
\vspace{-0.3em}

\textbf{Performance} \textbf{Analysis} \textbf{Among LLMs.} 
As for comparison among LLMs, DeepSeek emerges as the top performer, achieving a precision score of 93.20\%--97.96\%, a recall score of 96.00\%, an accuracy score of 94.50\%--97.00\%, and a F1 score of 94.58\%--96.97\% across both benchmarks. In contrast, Claude's performance is notably weaker, especially on the side of $\text{FixtureEval}_{\text{G}}$ with almost all metrics lower than 90.00\%, except precision. We dig deeper into each misclassified sample, and found that Claude often attempts to circumvent the single-line constraint through ``code compression" tactics. A clear example is the function \verb|get_dataclass_fields| shown in Figure \ref{fig:single & multi}, which requires a dataclass instance as input. Claude then generates a single line \verb|get_dataclass_fields(da| \verb|taclass(type('TestClass', (), {'x': int, 'y': str}))())|. It nests the entire dynamic class creation and instantiation process directly into the function invocation. Although it is an executable invocation in a single line, it does not follow the developers' programming conventions and damages the code readability. Thus, Claude only obtains a relatively high precision but obtains a lot of false negative predictions, i.e., relatively low recall scores.
Such clever but misguided behavior embeds the environmental setup, but undermines the premise of IBC: fixture-dependent samples cannot be invoked in a single code line, uncovering one of the weaknesses of IBC that could be improved in the future. However, considering that IBC overall consistently improves the fixture-dependence classification performance of all LLMs across all metrics, it is still empirically effective.
\begin{boxK}
\small \faIcon{pencil-alt} \textbf{Finding 8:} 
DeepSeek overall performs the best among all LLMs under test, while Claude performs the worst, mainly owing to its ``code compression'' tactics that bypass the premise of our IBC.  
\end{boxK}
\vspace{-0.3em}

\begin{table}[htbp]
\centering
\caption{Performance of \textsc{Fixturize} Combined With Existing Tools}
\small
\resizebox{\linewidth}{!}{
\begin{tabular}{lcccccc} 
\toprule
\multirow{2}{*}{\textbf{Approach}} & \multicolumn{3}{c}{\textbf{$\text{FixtureEval}_{\text{L}}$} } 
& \multicolumn{3}{c}{\textbf{$\text{FixtureEval}_{\text{G}}$}} \\
\cmidrule(lr){2-4} \cmidrule(lr){5-7}
& \textbf{SuitePS}& \textbf{LCov}& \textbf{BCov}& \textbf{SuitePS}& \textbf{LCov}& \textbf{BCov}\\
\midrule
C\textsc{over}U\textsc{p}& 65.50\% (45.00\%) & 47.14\% (41.49\%) & 45.19\% (37.99\%)
& 57.00\% (44.76\%)& 42.71\% (32.67\%)& 36.49\% (30.64\%)\\ 
C\textsc{over}U\textsc{p}+\textsc{Fixturize}& \textbf{75.50\%} (\textbf{65.00\%})& \textbf{60.85\%} (\textbf{63.63\%})& \textbf{64.29\%} (\textbf{66.59\%})& \textbf{58.00\%} (\textbf{46.67\%})& \textbf{44.69\%} (\textbf{36.25\%})& \textbf{38.64\%} (\textbf{34.42\%})\\ [-1pt] 
 Improvement Rate &\colorbox{light cyan}{15.27\% (44.44\%)}& 29.08\% (53.36\%) & 42.27\% (75.28\%) &\colorbox{light cyan}{1.75\% (4.26\%)}& 4.61\% (10.96\%)& 5.89\% (12.33\%)\\[-1pt] 
\midrule
\textsc{Pynguin}& 77.50\% (80.00\%)& 45.51\% (39.50\%) & 27.11\% (19.21\%) & 48.50\% (48.57\%)& 19.93\% (15.32\%)& 8.76\% (4.35\%)\\ 
\textsc{Pynguin}+\textsc{Fixturize}& 70.00\% (65.00\%)& 60.45\% (63.63\%) & 58.75\% (66.59\%) & 47.50\% (46.67\%)& 25.95\% (36.25\%)& 19.51\% (34.42\%)\\ [-1pt] 
Improvement Rate & -9.68\% (-18.75\%) &\colorbox{light cyan}{32.83\% (61.09\%)}&\colorbox{light cyan}{116.71\% (246.64\%)}& -2.06\% (-3.92\%)& \colorbox{light cyan}{30.24\% (136.70\%)}& \colorbox{light cyan}{122.60\% (691.65\%)}\\[-1pt] 
\bottomrule
\multicolumn{7}{p{18cm}}{\textit{Note:} Values in parentheses represent performance on the subset of fixture-dependent samples. Besides, we use \colorbox{light cyan}{light cyan} background in table cells to denote the highest improvement and \textbf{bold} values to highlight the highest rate in terms of SuitePS/LCov/BCov.}
\end{tabular}
}
\label{tab:combined_by_method}
\end{table}

\subsection{RQ4: Integration with Existing Approaches} 
To answer RQ4, we assess the extent to which \textsc{Fixturize} can enhance existing SOTA test generation tools on FixtureEval. Specifically, we select two representative testing tools, i.e., an LLM-based C\textsc{over}U\textsc{p} and a search-based \textsc{Pynguin}. Subsequnetly, we integrate them into \textsc{Fixturize} to handle those fixture-independent samples categorized by IBC.
Before examining the results, we highlight an important detail in C\textsc{over}U\textsc{p}’s workflow. By design, C\textsc{over}U\textsc{p} only retains a generated test suite if every test case within it passes. Thus, we cannot acurrately evaluate its PR, EX, and CasePS. 
Since both C\textsc{over}U\textsc{p} and \textsc{Pynguin} are coverage-oriented systems, we limit our comparison to the most direct and meaningful metrics: SuitePS, Line Coverage (LCov), and Branch Coverage (BCov). To ensure fairness, we also align our setup with C\textsc{over}U\textsc{p} by using GPT-4o as the underlying LLM throughout the experiments. 

As shown in Table \ref{tab:combined_by_method}, the integration with C\textsc{over}U\textsc{p} demonstrates a clear and consistent improvement across all metrics. On $\text{FixtureEval}_{\text{L}}$, \textsc{Fixturize} improves the SuitePS of C\textsc{over}U\textsc{p} from 65.50\% to 75.50\%. More significantly, owing to more passing test suites, it boosts LCov from 47.14\% to 60.85\% and BCov from 45.19\% to 64.29\%. When focusing exclusively on the fixture-dependent subset (parenthesized values), the impact of \textsc{Fixturize} becomes more evident, with SuitePS improving by 44.44\% , LCov by 53.36\%, and BCov by 75.28\%. Similar gains are observed on the more challenging $\text{FixtureEval}_{\text{G}}$ dataset. These results strongly indicate that the proactive fixture analysis of \textsc{Fixturize} succesfully resolved a series of test generation tasks with complex environment setup that C\textsc{over}U\textsc{p} previously cannot handle, thereby improving correctness and coverage.


    
    
    



\begin{figure}
    \centering
    \includegraphics[width=1\linewidth]{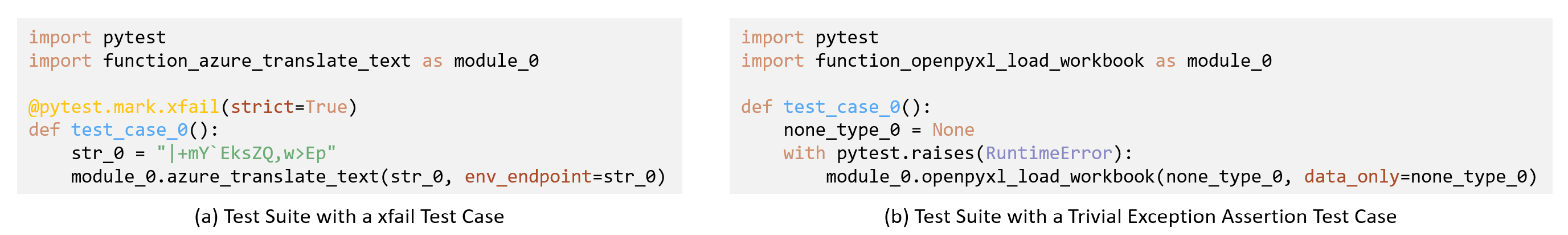}
    \caption{Examples of Low-Quality Test Suites Generated by Pynguin}
    \label{fig:pynguin-exmp}
\end{figure}

The results for \textsc{Pynguin} present a more complicated and, frankly, more interesting situation. While the integration of \textsc{Fixturize} leads to a substantial increase in both line and branch coverage, SuitePS shows a decrease, dropping from 48.50\%--77.50\% to 47.50\%--70.00\% across both benchmarks. However, based on a detailed analysis, we found that 68.2\% of \textsc{Pynguin}-generated test cases on fixture-dependent samples are either decorated with \texttt{xfail} or wrapped with deliberately-thrown exceptions as shown in Figure \ref{fig:pynguin-exmp}.
This phenomenon is an inherent byproduct of \textsc{Pynguin}’s generation strategy: since its evolutionary algorithm (e.g., DynaMOSA) prioritizes code coverage as the primary fitness function, it often retains crashing inputs that explore previously uncovered code paths despite being semantically invalid. Furthermore, due to the absence of environmental contexts and precise type information in dynamic Python environments, \textsc{Pynguin} frequently resorts to random type guessing; when it fails to satisfy complex object dependencies or environmental setups (fixtures) within its search budget, it exports these "half-baked" cases as \texttt{xfail} or exception-wrapped tests to maintain the integrity of the resulting suite. Consequently, while these tests contribute to an inflated SuitePS by bypassing core logic, they lack actual execution validity and fail to represent meaningful test scenarios.
In contrast, \textsc{Fixturize} guides the generation process towards creating more comprehensive, multi-scenario tests that proactively attempt to construct the required fixtures. This, on one hand, raises the complexity of the generated tests, naturally leading to a lower SuitePS. But it also makes \textsc{Fixturize} superior against \textsc{Pynguin} in terms of both line and branch coverages by 30.24\%--32.83\% and 116.71\%--122.60\%, respectively. If we focus exclusively on fixture-dependent samples, the advantage of \textsc{Fixturize} on test coverage is much more prominent with 691.65\% at most. This demonstrates that \textsc{Fixturize} makes \textsc{Pynguin} generate tests of higher quality and greater practical value.


\vspace{-0.3em}
\begin{boxK}
\small \faIcon{pencil-alt} \textbf{Finding 9:} 
\textsc{Fixturize} significantly enhances the practical performance of existing testing tools by resolving their fixture dependence, which is difficult to handle.  
\end{boxK}
\vspace{-0.3em}


\begin{table}[htbp]
\centering
\caption{Classification Performance on $\text{FixtureEval}_{\text{J}}$ Benchmark}
\setlength{\tabcolsep}{13pt}
\scriptsize
\begin{tabular}{llcccc} 
\toprule
\textbf{LLM} & \textbf{Approach} & \textbf{Precision} & \textbf{Recall} & \textbf{Accuracy} & \textbf{F1 Score} \\
\midrule
\multirow{3}{*}{DeepSeek} 
& Direct LLM classification & 61.25\% & 98.00\% & 68.00\% & 75.38\% \\
& Invocation-Based Classification & 94.34\% & \textbf{100.00\%} & 97.00\% & 97.09\% \\[-2pt] 
&Improvement Rate & \colorbox{light cyan}{54.02\%} & 2.04\% & \colorbox{light cyan}{42.65\%} & \colorbox{light cyan}{28.80\%} \\[-1pt] 
\midrule
\multirow{3}{*}{GPT-4o} 
& Direct LLM classification & 67.59\% & 98.00\% & 75.50\% & 80.00\% \\
& Invocation-Based Classification & 90.09\% & \textbf{100.00\%} & 94.50\% & 94.79\% \\
&Improvement Rate & 33.29\% & 2.04\% & 25.17\% & 18.49\% \\
\midrule
\multirow{3}{*}{Claude} 
& Direct LLM classification & 69.92\% & 93.00\% & 76.50\% & 79.83\% \\
& Invocation-Based Classification & 94.23\% & 98.00\% & 96.00\% & 96.08\% \\
&Improvement Rate & 34.77\% & 5.38\% & 25.49\% &20.36\% \\
\midrule
\multirow{3}{*}{Qwen} 
& Direct LLM classification & 69.17\% & 92.00\% & 75.50\% & 78.97\% \\
& Invocation-Based Classification & \textbf{96.12\%} & 99.00\% & \textbf{97.50\%} & \textbf{97.54\%} \\[-2pt] 
& Improvement Rate & 38.96\% & \colorbox{light cyan}{7.61\%} & 29.14\% & 23.52\% \\[-1pt] 
\bottomrule
\multicolumn{6}{p{13cm}}{\textit{Note:} We use \colorbox{light cyan}{light cyan} background in table cells to denote the highest improvement and \textbf{bold} values to highlight the highest rate in terms of Precision/Recall/Accuracy/F1 Score.}
\end{tabular}
\label{tab:classifier_performance}
\end{table}

\subsection{RQ5: Cross-Language Generalization}
\label{RQ5}
Based on our previously constructed $\text{FixtureEval}_{\text{J}}$, we first examine the classification performance of \textsc{Fixturize}, as its accuracy is fundamental to the entire workflow. As shown in Table \ref{tab:classifier_performance}, IBC remains remarkably effective in the Java ecosystem, where it significantly outperforms direct LLM classification across all LLMs and metrics. For instance, IBC achieves F1-scores consistently above 94\% and near-perfect recall for all LLMs. As for precision and accuracy, IBC also reaches quite high scores of 93.70\% and 96.25\% on average, respectively. Among different LLMs, DeepSeek obtains the most substantial gain on most metrics. For example, IBC improves its precision by 54.02\%, accuracy by 42.65\%, and F1 score by 28.80\%. These results confirm that our IBC strategy is language-agnostic in identifying fixture-dependent code samples.

\begin{table}[htbp]
\scriptsize
\centering 
\caption{Test Generation Performance on $\text{FixtureEval}_{\text{J}}$ Benchmark}
\setlength{\tabcolsep}{9pt} 
\resizebox{\textwidth}{!}{
    \begin{tabular}{llcccc}
    \toprule
    \textbf{LLM}& \textbf{Approach}& \textbf{PR}& \textbf{EX} & \textbf{CasePS}& \textbf{SuitePS}\\
    \midrule
    \multirow{3}{*}{DeepSeek} & Direct LLM prompting & 47.00\% (20.75\%) & 54.49\% (20.22\%) & 42.44\% (8.84\%) & 26.50\% (9.43\%) \\
    & \textsc{Fixturize} & 65.50\% (55.66\%) & 70.34\% (55.54\%) & 46.52\% (17.79\%) & 30.00\% (16.04\%) \\[-2pt] 
    & Improvement Rate & \colorbox{light cyan}{39.36\% (168.24\%)}& 29.09\%\colorbox{light cyan}{(174.68\%)} & 9.61\%\colorbox{light cyan}{(101.24\%)} & 13.21\% (70.10\%) \\[-2pt] 
    \midrule
    \multirow{3}{*}{GPT-4o} & Direct LLM prompting & 59.00\% (42.34\%) & 59.02\% (42.55\%) & 36.39\% (12.75\%) & 24.50\% (6.31\%) \\
    & \textsc{Fixturize} & \textbf{77.50\%} (\textbf{75.68\%}) & \textbf{77.48\%} (\textbf{75.77\%})& 40.24\% (19.53\%) & 29.50\% (15.32\%) \\[-2pt] 
    & Improvement Rate & 31.36\% (78.74\%) & \colorbox{light cyan}{31.28\%} (78.07\%)& 10.58\% (53.18\%) &  20.41\%\colorbox{light cyan}{(142.79\%)} \\[-2pt] 
    \midrule
    \multirow{3}{*}{Claude} & Direct LLM prompting & 66.00\% (51.92\%) & 65.81\% (51.71\%) & 50.30\% (30.99\%) & 47.00\% (38.46\%) \\
    & \textsc{Fixturize} & 77.00\% (73.08\%) & 77.05\% (73.17\%) & \textbf{57.62\%} (\textbf{44.79\%}) &\textbf{59.50\%} (\textbf{62.50\%})\\[-2pt] 
    & Improvement Rate & 16.67\% (40.76\%) &17.08\% (41.50\%) & \colorbox{light cyan}{14.55\%} (44.53\%)& \colorbox{light cyan}{26.60\%} (62.51\%) \\[-2pt] 
    \midrule
    \multirow{3}{*}{Qwen} & Direct LLM prompting & 39.00\% (15.53\%) & 39.98\% (15.69\%) & 30.11\% (6.47\%) & 21.00\% (5.83\%) \\
    & \textsc{Fixturize} & 47.50\% (32.04\%) & 48.52\% (32.16\%) & 32.55\% (11.18\%) & 23.50\% (10.68\%) \\
    & Improvement Rate & 21.79\% (106.31\%) & 21.36\% (104.97\%) & 8.10\% (72.80\%) & 11.90\% (83.19\%) \\
    \midrule
    \multirow{3}{*}{Average} & Direct LLM prompting & 52.75\% (32.64\%) & 54.83\% (32.54\%) & 39.81\% (14.76\%) & 29.75\% (15.01\%) \\
    & \textsc{Fixturize} & 66.88\% (59.12\%) & 68.35\% (59.16\%) & 44.23\% (23.32\%) & 35.63\% (26.14\%) \\
    & Improvement Rate & 27.30\% (98.51\%) & 24.70\% (99.81\%) & 10.71\% (67.94\%) & 18.03\% (89.65\%) \\
    \bottomrule
    \multicolumn{6}{p{13cm}}{\textit{Note:} Values in parentheses represent performance on the subset of fixture-dependent samples. Besides, we use \colorbox{light cyan}{light cyan} background in table cells to denote the highest improvement and \textbf{bold} values to highlight the highest rate in terms of PR/EX/CasePS/SuitePS.}
    \end{tabular}
} 
\label{tab:Java_test_performance}
\end{table}

With this reliable classification as a foundation, we now turn to the overall test generation performance shown in Table \ref{tab:Java_test_performance}. Compared with Python, generating tests for Java programs, especially for those fixture-dependent ones, is significantly harder for LLMs, with the average baseline SuitePS standing at a mere 29.75\% on average. However, \textsc{Fixturize} still provides a powerful boost. The most dramatic improvements are in syntactic validity and executability, with PR rising from 52.75\% to 66.88\% and EX from 54.83\% to 68.35\% on average. This indicates that a primary failure for the direct LLM prompting is the generation of code with syntax or dependency errors. Our framework effectively alleviates this, transforming a large volume of unusable output into compilable and runnable tests. This benefit is especially apparent for LLMs as evidenced by DeepSeek, whose PR jumps from 47.00\% to 65.50\% and EX from 54.49\% to 70.34\%. 
As for CasePS and SuitePS, although the improvements are relatively modest, LLMs still obtain a considerable performance gain of 10.71\% and 18.03\% in terms of each metric on average, confirming that \textsc{Fixturize} moves beyond merely making tests runnable to improving its coverage performance.
The impact of \textsc{Fixturize} becomes clearer when it comes to the fixture-dependent subset (parenthesized values). To be specific, \textsc{Fixturize} on average improves the PR score by 98.51\%, EX score by 99.81\%, CasePS by 67.94\%, and SuitePS by 89.65\%. Overall, the liftings on PR and EX are still more significant than those on CasePS and SuitePS, underscoring that generating a functionally correct test suite remains a more difficult challenge than simply producing executable code.  



 \vspace{-0.3em}
\begin{boxK}
\small \faIcon{pencil-alt} \textbf{Finding 10:} 
Our evaluation on $\text{FixtureEval}_{\text{J}}$ confirms the strong generality of \textsc{Fixturize} on fixture-dependence classification and associated test generation, showing that \textsc{Fixturize} addresses a core, language-agnostic challenge in test generation rather than relying on Python-specific patterns.
\end{boxK}
\vspace{-0.3em}

\section{Threats to Validity}
\textbf{Threats to external validity} concern the generalizability of our findings. Our study relies on the FixtureEval dataset and a specific set of four SOTA LLMs (DeepSeek, GPT-4o, Claude, and Qwen). While our primary evaluation (RQ1-RQ4) focuses on Python, we explicitly mitigate the language-generalization threat in RQ5 by evaluating on our \textbf{$\mathrm{FixtureEval}_{\mathrm{J}}$} dataset. Besides, it is important to note that $\mathrm{FixtureEval}_{\mathrm{J}}$ consists of samples from four projects selected from C\textsc{hat}T\textsc{ester}'s empirical study, which may exhibit an uneven distribution of software categories. However, we consider this threat to be minimal, as the generality of \textsc{Fixturize} on diverse software categories has been examined on the Python side and $\mathrm{FixtureEval}_{\mathrm{J}}$ is used for cross-PL investigation only.
A second major threat, data contamination, is directly and rigorously safeguarded against by evaluating all methods on our \textbf{$\mathrm{FixtureEval}_{\text{L}}$} subset. A third external threat relates to our choice of LLMs. Our framework relies on powerful SOTA LLMs, as we observed in preliminary experiments that smaller, open-source LLMs cannot often understand the complex instructions and contextual nuances required for structured fixture generation. Thus, our results may not generalize to these less capable LLMs.

\noindent\textbf{Threats to internal validity} concern potential biases in our experimental design and implementation. To ensure fair and comprehensive comparisons, we deliberately selected baseline approaches that represent distinct technical paradigms in test case generation. Specifically, we compared \textsc{Fixturize} against C\textsc{over}U\textsc{p}, a SOTA LLM-based approach that leverages the generative power of LLMs, and \textsc{Pynguin}, a widely-used search-based technique that relies on evolutionary algorithms to generate tests. This selection ensures that our evaluation covers the two dominant and contrasting methodologies in the field. We strictly adhered to the official configurations and recommendations for all baselines to prevent any unfair advantage or implementation bias. Furthermore, to mitigate potential errors in our own implementation of \textsc{Fixturize}, we have conducted rigorous testing and publicly released our complete replication package, which includes all source code, datasets, and experimental setups. This commitment to transparency allows for the verification of our findings and promotes reproducibility in future research.

\noindent\textbf{Threats to construct validity} relate to whether our measurements adequately reflect the concept of test quality in fixture-aware test generation. To mitigate this, we adopt an evaluation design that corresponds to the two core tasks of our approach. For the classification stage (RQ1), we apply standard metrics including precision, accuracy, recall, and F1-score, which together provide a balanced assessment of the LLM’s ability to distinguish fixture-dependent from fixture-independent functions. For test generation (RQ2 and RQ3), we employ both static and dynamic evaluation. Syntactic validity is examined through successful parse rate (PR), while execution robustness is measured by successful execution rate (EX). Test correctness is evaluated at both the individual case level (CasePS) and the entire suite level (SuitePS), providing a more comprehensive view of test utility and reliability. Coverage measures (LCov and BCov) used in RQ4 follow the protocol of involved baselines but do not serve as the only indicators of effectiveness, as coverage alone may not correspond directly to practical correctness. Based on the above, the related threat can be minimized. 

\section{Conclusion and Future Work}

This work contributed to the first LLM fixture-aware unit test generation approach, named \textsc{Fixturize}, and established the first dedicated benchmark, FixtureEval, for evaluating fixture generation capabilities. We conducted extensive experiments with SOTA LLMs, comparing \textsc{Fixturize} against traditional approaches and common LLM-based approaches. Results demonstrate that \textsc{Fixturize} significantly outperforms competitive baselines, especially on functions requiring complex fixture setup, showing the motivation and necessity of incorporating structured fixture awareness into LLM test generation. Afterwards, we further analyzed the efficacy of \textsc{Fixturize}'s different components, namely IBC, EIC, and UTG, offering valuable guidance for designing robust fixture-aware test generators. Finally, we enhanced existing tools by integrating \textsc{Fixturize} with them, and evaluated the resulting performance in cross-language settings to validate its generality.   Our findings provide practical insights to facilitate further research and application. In the future, we will incorporate retrieval-augmented generation into \textsc{Fixturize} to further advance its practical classification and generation performance, as well as extend FixtureEval to include a broader range of PL for more comprehensive assessments.


\section*{Acknowledgments}

This work was partially supported by the National Natural Science Foundation of China (Grant Nos. 62502283 and U24B20149), the Natural Science Foundation of Shandong Province (Grant No. ZR2024QF093), the Young Talent of Lifting Engineering for Science and Technology in Shandong, China (Grant No. SDAST2025QTB031).

\bibliographystyle{ACM-Reference-Format}
\bibliography{sample-base}


\end{document}